\documentclass[aps,prd,reprint,amsmath,amssymb,superscriptaddress,floatfix]{revtex4}
\usepackage{graphicx}
\usepackage{amsfonts}
\usepackage{amssymb}
\usepackage{rotating}
\usepackage{booktabs}
\usepackage{xcolor}
\usepackage{soul}
\usepackage{color}
\usepackage{slashed}
\usepackage{multirow}
\usepackage{makecell}
\usepackage{epsf}
\usepackage{ulem}
\usepackage{cancel}
\usepackage{color,bm}
\usepackage[colorinlistoftodos]{todonotes}

\usepackage[colorlinks=true,citecolor=cyan,urlcolor=blue,bookmarks=true,bookmarks=true,bookmarksopen=true,bookmarksnumbered=true,bookmarksopenlevel=3]{hyperref}

\definecolor{airforceblue}{rgb}{0.36, 0.54, 0.66}
\definecolor{steelblue}{rgb}{0.27, 0.51, 0.71}
\definecolor{amber}{rgb}{1.0, 0.49, 0.0}

\newcommand{\s}{\slashed}

\begin{document}

\title{Single spin asymmetry $A_{UL}^{\sin(\phi_h-\phi_R)}$ in dihadron semi-inclusive DIS}
\author{Xuan Luo}
\author{\textsc{Hao Sun}\footnote{Corresponding author: haosun@mail.ustc.edu.cn \hspace{0.2cm} haosun@dlut.edu.cn}}
\author{\textsc{Yi-Ling Xie}}
\affiliation{Institute of Theoretical Physics, School of Physics, Dalian University of Technology, \\ No.2 Linggong Road, Dalian, Liaoning, 116024, P.R.China }
\date{\today}

\begin{abstract}

In this paper we study the single longitudinal spin asymmetry $A_{UL}^{\sin(\phi_h-\phi_R)}$ of dihadron production 
in semi-inclusive deep inelastic scattering (SIDIS) via helicity-dependent dihadron fragmentation function (DiFF), which describes the correlation of the longitudinal polarization of a fragmenting quark with the transverse momenta of the produced hadron pair. 
Recently experimental searching for this azimuthal asymmetry in dihadron SIDIS by the COMPASS Collaboration yielded a very small signal.
Here we calculate this unknown T-odd DiFF $G_1^\perp$ using a spectator model to access the asymmetry and clarify why the signal is very small.
The transverse momentum dependent (TMD) factorization method, in which the transverse momentum of the final state hadron pair leaves unintegrated, has been applied.
We estimate the $\sin(\phi_h-\phi_R)$ asymmetry at the kinematics of COMPASS experiments and compare with the data.
What's more, the predictions on the same asymmetry are also made at the Electron Ion Collider (EIC).  

\vspace{0.5cm}
\end{abstract}
\maketitle
\setcounter{footnote}{0}

\section{INTRODUCTION}
\label{I}

The study of the dihadron fragmentation functions (DiFFs) describing the probability 
that a quark hadronizes into two hadrons is of great interest both in theory and in experiment.
The DiFFs were firstly proposed in Ref. \cite{Konishi:1979cb}. 
Their evolution equations have been studied in Ref.\cite{Vendramin:1981gi,Vendramin:1981te}, 
and later in Ref.\cite{Ceccopieri:2007ip} where the evolution was discussed, for the first time, as functions of the hadron pair invariant mass $M_h$.
Ref.\cite{Collins:1994ax} introduced the transversely polarized DiFF as spin analyzer of the
transversely polarized fragmenting quark, which later lead to the definition of $H_1^\sphericalangle$.
The first comprehensive study of dihadron fragmentation has been presented in Ref.\cite{Bianconi:1999cd} up to leading twist, where the relevant DiFFs have been defined.
DiFFs have been used to investigate the transverse spin phenomena of the nucleon, 
and they can act as analyzers of the spin of the fragmenting quark \cite{Collins:1993kq,Jaffe:1997hf}.
The authors in Ref.\cite{Radici:2001na} introduced the method of partial-wave analysis, which makes the connection between two-hadron production and spin-one production clear. 
Then in Ref.\cite{Bacchetta:2003vn} the analysis was extended to subleading twist integrated over the transverse component of the momentum of the hadron pair. This analysis can be directly promoted into transverse momentum dependent case. The cross section expression for production of two hadrons in SIDIS within TMD factorization is presented in Ref.\cite{Gliske:2014wba}.
Physicists started to focus on these functions by searching for a mechanism 
to extract the chiral-odd transversity distribution in an alternative and technically simpler way than the Collins effect \cite{Collins:1992kk}.
Transversity was extracted for the first time from data on single hadron SIDIS by a convolution $h_1 \otimes H_1^\perp$. One must involves the transverse momenta of quarks working for the chiral-odd Collins fragmentation function. An alternative method to access the transversity PDF
requires only standard collinear factorization. In this mechanism, the chiral-odd DiFF $H_1^\sphericalangle$ \cite{Radici:2001na,Bacchetta:2002ux} 
plays an important role in accessing transversity distribution $h_1$, as it couples with $h_1$ at the leading-twist level in the collinear factorization.
The azimuthal asymmetry in the distribution of charged pion pairs in $e^+ e^-$ annihilation was measured by the BELLE collaboration \cite{Vossen:2011fk}, opening the way to the first parametrization of $H_1^\sphericalangle$ for the up and down quarks \cite{Courtoy:2012ry}.
In Ref. \cite{Bacchetta:2011ip,Bacchetta:2012ty,Radici:2015mwa,Radici:2016lam,Radici:2018iag}, the authors extracted $h_1$ from SIDIS and proton-proton collision data.
At the same time, the model predictions of the DiFFs were performed by the spectator model \cite{Bianconi:1999uc,Bacchetta:2006un,Bacchetta:2008wb} 
and by the Nambu-Jona-Lasinio (NJL) quark model \cite{Matevosyan:2013aka,Matevosyan:2013eia,Matevosyan:2017alv,Matevosyan:2017uls}.

The HERMES collaboration \cite{Airapetian:2008sk} and COMPASS collaboration \cite{Adolph:2012nw,Adolph:2014fjw} 
have measured the azimuthal asymmetries of the SIDIS process in which hadron pair are produced with an unpolarized target or a transversely polarized target.
The azimuthal angular dependences involving dihadron fragmentation in the leading order cross section of dihadron SIDIS, were presented in Ref. \cite{Radici:2001na}.
Then in Ref. \cite{Bacchetta:2003vn}, the authors gived a complete list of cross section and spin asymmetries up to the subleading twist.
Very recently, results on the azimuthal asymmetries in dihadron production with the longitudinally polarized proton target 
were also obtained by the COMPASS collaboration \cite{Sirtl:2017rhi}.
Considering the dihadron cross section in a transverse momentum dependent (TMD) factorization approach 
with the incident lepton beam being unpolarized or longitudinally polarized, a $\sin(\phi_h-\phi_R)$ modulation, 
among many modulations, comes into our view. Here $\phi_h$ denotes the azimuthal angle of the hadron pair system 
and $\phi_R$ is the angle between the lepton plane and two-hadron plane. Within the TMD factorization appoach, 
the dihadron SIDIS cross section is written as a convolution of TMD distribution functions and TMD-DiFFs.
TMD factorization extends the collinear factorization by accounting for the parton transverse momentum. On the experimental side, the COMPASS measurements show that the $\sin(\phi_h-\phi_R)$ asymmetry is compatible with zero whithin experimental precision. 
On the theoretical point of view, within the parton model, there are two sources contributing to the $\sin(\phi_h-\phi_R)$ asymmetry, 
namely the coupling of the twist-2 distribution $g_{1L}$ and the T-odd DiFF $G_1^\perp$.

In this paper we investigate the $\sin(\phi_h-\phi_R)$ asymmetry by using the relevant parton distribution functions (PDFs) and DiFFs based on the spectator model.
After performing partial waves expansion, one possible contribution of this asymmetry is $g_{1L} G_{1,OT}^\perp$ 
where $G_{1,OT}^\perp$ stems from interference of $s$- and $p$-waves. The other contribution comes from $g_{1L} G_{1,LT}^\perp$ 
where $G_{1,LT}^\perp$ originates from the interference of two $p$-waves with different polarizations. 
The latter contribution will vanish after performing the polar angle integration. 
We adopt the spectator model \cite{Bacchetta:2006un} to calculate $G_{1,OT}^\perp$ 
and find that, to obtain a nonvanishing $G_{1,OT}^\perp$, one must consider loop contributions. 
Applying the spectator model results for the PDFs and DiFFs, we estimate the $\sin(\phi_h-\phi_R)$ asymmetry 
at the COMPASS kinematics and compare it with the COMPASS preliminary data. 

The paper is organized as follows. In Sec.\ref{II} we review the basic framework of DiFFs. 
We apply the spectator model to calculate the T-odd DiFF $G_{1,OT}^\perp$ in Sec.\ref{III}. 
In Sec.\ref{IV}, we present the theoretical framework of the $\sin(\phi_h-\phi_R)$ azimuthal asymmetry 
in dihadron SIDIS with unpolarized lepton beam scattering off a longitudinally polarized proton target. 
Then in Sec.\ref{V}, we give the numerical results of the $\sin(\phi_h-\phi_R)$ azimuthal asymmetry at the kinematics of COMPASS measurements 
and also make the predictions at the EIC. We finally summarize our work in Sec.\ref{VI}.

\section{Basic framework of dihadron fragmentation function}
\label{II}

We start from the physical picture $q \to \pi^+ \pi^- X$ that a quark with momentum $k$ fragments into two leading unpolarized pions with mass $M_1, M_2$, 
and momenta $P_1, P_2$ respectively. The invariant mass of the hadron pair is defined by $M_h$. It is useful to interpose two vectors 
$P_h=P_1+P_2$ and $R=(P_1-P_2)/2$. We present an arbitrary four vector $a$ in the light-cone coordinates 
$\displaystyle a^\pm = (a^0 \pm a^3)/\sqrt{2}$ and $\vec{a}_T=(a^1,a^2)$. This provide all components as $[a^-,a^+,\vec{a}_T]$.
We also introduce $z_i$ denoting the light-cone fraction of the fragmenting quark momentum carried by hadron $h_i$. 
The light-cone fraction of the fragmenting quark momentum carried by the final hadron pair is defined by $z$.

It is convenient to choose the $\hat{z}$ axis according to the condition $\vec{P}_{hT}=0$. 
Therefore, the momenta $P_h^\mu$, $k^\mu$ and $R^\mu$ can be written as \cite{Bacchetta:2006un}
\begin{eqnarray} \label{eq:1}
\begin{aligned}
P_h^{\mu}&=\left[P^-_h,\frac{M_h^2}{2P^-_h}, \vec{0}_T \right] \\ 
k^{\mu}  &=\left[\frac{P_h^-}{z},\frac{z(k^2+\vec{k}_T^2)}{2P_h^-},\vec{k}_T \right] \\
R^{\mu}  &=\left[-\frac{|\vec{R}|P^-_h}{M_h}\cos\theta,\frac{|\vec{R}|M_h}{2P^-_h}\cos\theta,|\vec{R}|\sin\theta\cos\phi_R,|\vec{R}|\sin\theta\sin\phi_R \right] \\
&=\left[-\frac{|\vec{R}|P^-_h}{M_h}\cos\theta,\frac{|\vec{R}|M_h}{2P^-_h}\cos\theta,  \vec{R}_T^x, \vec{R}_T^y \right] ,
\end{aligned}
\end{eqnarray}
where
\begin{eqnarray} \label{eq:2}
\begin{aligned}
|\vec{R}| = \sqrt{\frac{M_h^2}{4}-m_\pi^2} \ \ .
\end{aligned}
\end{eqnarray}
Here $m_\pi$ is the mass of pion. It is desired to notice that in order to perform partial-wave expansion, 
we have reformulated the kinematics in the center of mass frame of the dihadron system. 
$\theta$ is the center of mass polar angle of the pair with respect to the direction of $P_h$ in the target rest frame \cite{Bacchetta:2002ux}. 
We can find some useful relations as
\begin{eqnarray} \label{eq:3}
\begin{aligned}
P_h \cdot R &= 0 \\
P_h \cdot k &= \frac{M_h^2}{2z}+z\frac{k^2+\vec{k}_T^2}{2} \\
R \cdot k &= \left( \frac{M_h}{2z}-z\frac{k^2+\vec{k}_T^2}{2M_h} \right) |\vec{R}| \cos\theta - \vec{k}_T \cdot \vec{R}_T \ \ .
\end{aligned}
\end{eqnarray}

\section{The model calculation of $G_{1,OT}^\perp$}
\label{III}

The TMD DiFFs $D_1$ and $G_1^\perp$ are extracted from the quark-quark correlator $\Delta(k, P_h, R)$
\begin{eqnarray} \label{eq:4}
\begin{aligned}
\Delta(k, P_h, R) \displaystyle &=\sum \kern -1.3 em \int_X \;
\int \frac{d^4 \xi}{(2\pi)^4} \; e^{\text{i} k\cdot\xi}\;
\langle 0|\psi(\xi) \,|P_h,R; X\rangle
\langle X; P_h,R|\, \overline{\psi}(0)\,|0\rangle
\Big|_{\xi^- = \vec{\xi}_T  =0} \; \\
&=\frac{1}{16\pi}\left\{ D_1 \slashed{n}_-+G_1^\perp \gamma_5 \frac{\varepsilon_T^{\rho\sigma} R_{T\rho} k_{T\sigma}}{M_h^2} \s{n}_- + \cdots \right\} .
\end{aligned}
\end{eqnarray}
Then we express the leading-twist quark-quark correlator Eq.(\ref{eq:4}) in terms of center of mass variables. The connection between the two representations is defined as
\begin{eqnarray} \label{eq:5}
\begin{aligned}
\Delta(z,k_T^2,\cos\theta,M_h^2,\phi_R) = \frac{|\vec{R}|}{16zM_h} \int dk^+ \Delta(k, P_h, R) .
\end{aligned}
\end{eqnarray}
By projecting out the usual Dirac structures, we obtain the following decomposition results
\begin{eqnarray} \label{eq:6}
\begin{aligned}
4 \pi \text{Tr}[\Delta(z,k_T^2,\cos\theta,M_h^2,\phi_R)\gamma^- \gamma^5] = \frac{\varepsilon_T^{\rho\sigma} R_{T\rho} k_{T\sigma}}{M_h^2}G_{1}^{\perp} ,
\end{aligned}
\end{eqnarray}
where $\gamma^-$ is the negative light-cone Dirac matrix.

The TMD DiFFs $D_1$, $G_1^\perp$ can be expanded in the relative partial waves of the dihadron system up to the $p$-wave level \cite{Bacchetta:2002ux}: 
\begin{eqnarray} \label{eq:7}
\begin{aligned}
D_{1}(z,k_T^2, \cos \theta,M_h^2)&=D_{1,OO}(z,M_h^2)+D_{1,OL}(z,M_h^2)\cos \theta+\frac{1}{4}D_{1,LL}(z,M_h^2)(3\cos^2\theta-1)
\\
&+\cos(\phi_k-\phi_R)\sin\theta(D_{1,OT}+D_{1,LT}\cos\theta) + \cos(2\phi_k-2\phi_R)\sin^2\theta D_{1,TT}, \\
G_1^\perp(z,k_T^2, \cos \theta,M_h^2)&=G_{1,OT}^\perp+G_{1,LT}^\perp \cos\theta + \cos(\phi_k-\phi_R) \sin\theta G_{1,TT}^\perp ,
\end{aligned}
\end{eqnarray}
where $G_{1,OT}^\perp$ comes from the interference of $s$- and $p$-waves, and $G_{1,LT}^\perp $ originates from the interference of two $p$-waves with different polarizations. $\phi_k$ is the azimuthal angle of quark transverse momentum $\vec{k}_T$ with respect to the lepton scattering plane. 

Since the contribution including $G_{1,LT}^\perp$ to the $\sin(\phi_h-\phi_R)$ asymmetry vanishes (see Sec.\ref{IV} in the following), 
we will work out the DiFF $G_{1,OT}^\perp$ in the spectator model below. 
In principle, the tree level correlator yields vanishing contribution to $G_{1,OT}^\perp$ because of the shortage of the imaginary phase. 
Whereas, the quark-dihadron interaction vertex $F^{s*}F^p$ required to be complex will generate this imaginary phase. 
Therefore, we can obtain the correlation function in a similar way as in Ref.\cite{Bacchetta:2006un}  
\begin{eqnarray} \label{eq:8}
\begin{aligned}
\Delta^q (k,P_h,R) &= \frac{1}{(2\pi)^4}\frac{(\s{k}+m)}{(k^2-m^2)^2}\left( F^{s*}e^{-\frac{k^2}{\Lambda_s^2}}
+F^{p*}e^{-\frac{k^2}{\Lambda_p^2}} \s{R} \right) (\s{k}-\s{P}_h+M_s) \left( F^{s}e^{-\frac{k^2}{\Lambda_s^2}}+F^{p}e^{-\frac{k^2}{\Lambda_p^2}} \s{R} \right) 
\\
& \times (\s{k}+m) \cdot 2\pi \delta((k-P_h)^2-M_s^2) ,
\end{aligned}
\end{eqnarray}
where $m$ and $M_s$ represent the masses of the fragmented quark and the spectator quark, respectively. 
The $s$-wave and $p$-wave vertex structures $F^s$ and $F^p$ are wirtten as \cite{Bacchetta:2006un}:
\begin{eqnarray} \label{eq:9}
\begin{aligned}
F^s&=f_s \\
F^p&=
f_\rho \frac{M_h^2-M_\rho^2-i\Gamma_\rho M_\rho}{(M_h^2-M_\rho^2)^2+\Gamma_\rho^2 M_\rho^2}
+f_\omega \frac{M_h^2-M_\omega^2-i\Gamma_\omega M_\omega}{(M_h^2-M_\omega^2)^2+\Gamma_\omega^2 M_\omega^2} 
- i f_\omega'\frac{\sqrt{\lambda(M_\omega^2,M_h^2,m_\pi^2)}\Theta(M_\omega-m_\pi-M_h)}{4\pi \Gamma_\omega 
M_\omega^2 [4M_\omega^2 m_\pi^2+\lambda(M_\omega^2,M_h^2,m_\pi^2)]^\frac{1}{4}} ,
\end{aligned}
\end{eqnarray}
where $\lambda(M_\omega^2,M_h^2,m_\pi^2)=[M_\omega^2-(M_h+m_\pi)^2][M_\omega^2-(M_h-m_\pi)^2]$ 
and $\Theta$ denotes the unit step function. The couplings $f_s$, $f_\rho$, $f_\omega$ and $f_\omega'$ are the model parameters. 
The first two terms of $F^p$ can be classified as the contributions of the $\rho$ and the $\omega$ resonances decaying into two pions. 
The masses and the widths of the two resonances can be accessed from the PDG \cite{Eidelman:2004wy}: 
$M_\rho=0.776$ GeV, $\Gamma_\rho=0.150$ GeV, $M_\omega=0.783$ GeV and $\Gamma_\omega=0.008$ GeV.

Putting Eq.(\ref{eq:8}) into Eq.(\ref{eq:5}), we acquire
\begin{eqnarray} \label{eq:10}
\begin{aligned}
\Delta^q (z,k_T^2,\cos\theta,M_h^2,\phi_R) &= \frac{|\vec{R}|}{256\pi^3 z(1-z) M_h k^-}\bigg[ 
|F^s|^2 e^{-\frac{2k^2}{\Lambda_{s}^2}}\frac{(\s{k}+m)(\s{k}-\s{P}_h+M_s)(\s{k}+m)}{(k^2-m^2)^2} 
\\
&+|F^p|^2 e^{-\frac{2k^2}{\Lambda_{p}^2}}\frac{(\s{k}+m)\s{R}(\s{k}-\s{P}_h+M_s)\s{R}(\s{k}+m)}{(k^2-m^2)^2} 
\\
&+F^{s*}F^p e^{-\frac{2k^2}{\Lambda_{sp}^2}}\frac{(\s{k}+m)(\s{k}-\s{P}_h+M_s)\s{R}(\s{k}+m)}{(k^2-m^2)^2} 
\\
&+ F^{s}F^{p*} e^{-\frac{2k^2}{\Lambda_{sp}^2}}\frac{(\s{k}+m)\s{R}(\s{k}-\s{P}_h+M_s)(\s{k}+m)}{(k^2-m^2)^2} \bigg] .
\end{aligned}
\end{eqnarray}
Here the $z$-dependent $\Lambda$-cutoffs $\Lambda_{sp}$ and $\Lambda_{s,p}$ have the relation
\begin{eqnarray} \label{eq:11}
\begin{aligned}
\frac{2}{\Lambda_{sp}^2}=\frac{1}{\Lambda_s^2}+\frac{1}{\Lambda_p^2} ,
\end{aligned}
\end{eqnarray}
where $\Lambda_{s,p}$ have the following structure :
\begin{eqnarray} \label{eq:12}
\begin{aligned}
\Lambda_{s,p}=\alpha_{s,p} z^{\beta_{s,p}}(1-z)^{\gamma_{s,p}} ,
\end{aligned}
\end{eqnarray}
and $\alpha$, $\beta$ and $\gamma$ are the parameters shown below. 
The $k^2$ term is fixed by the on-shell condition of the spectator
\begin{eqnarray} \label{eq:13}
\begin{aligned}
k^2=\frac{z}{1-z}\vec{k}_T^2+\frac{M_s^2}{1-z}+\frac{M_h^2}{z} .
\end{aligned}
\end{eqnarray}

In Eq.(\ref{eq:10}), the terms coupled with $|F^s|^2$ and $|F^p|^2$ characterize the pure $s$- and $p$-wave contributions, 
thus they will not make a difference in the interference of $s$ and $p$-waves functions $G_{1,OT}^\perp$. 
While the last two terms represent the $s$- and $p$-wave interference, and they do contribute to the $G_{1,OT}^\perp$ 
with the necessary imaginary phase originated from $F^p$. Then we obtain the tree-level result for $G_{1,OT}^\perp$
\begin{eqnarray} \label{eq:14}
\begin{aligned}
&G_{1,OT}^\perp=\frac{1}{4\pi^2}\frac{M_h |\vec{R}| m}{1-z}e^{-\frac{2k^2}{\Lambda_{sp}^2}} \frac{1}{(k^2-m^2)^2}.
\end{aligned}
\end{eqnarray}
In this paper we choose the input quark mass to be zero GeV, which is consistent with Ref. \cite{Bacchetta:2006un}. 
We have checked that a small quark mass value mainly don't affect the results of DiFF and corresponding asymmetry. 
Thus, the tree level $G_{1,OT}^\perp$ vanishes and we have to consider the loop contributions for this DiFF.

\begin{figure}[htp]
\centering
\includegraphics[height=3cm,width=4cm]{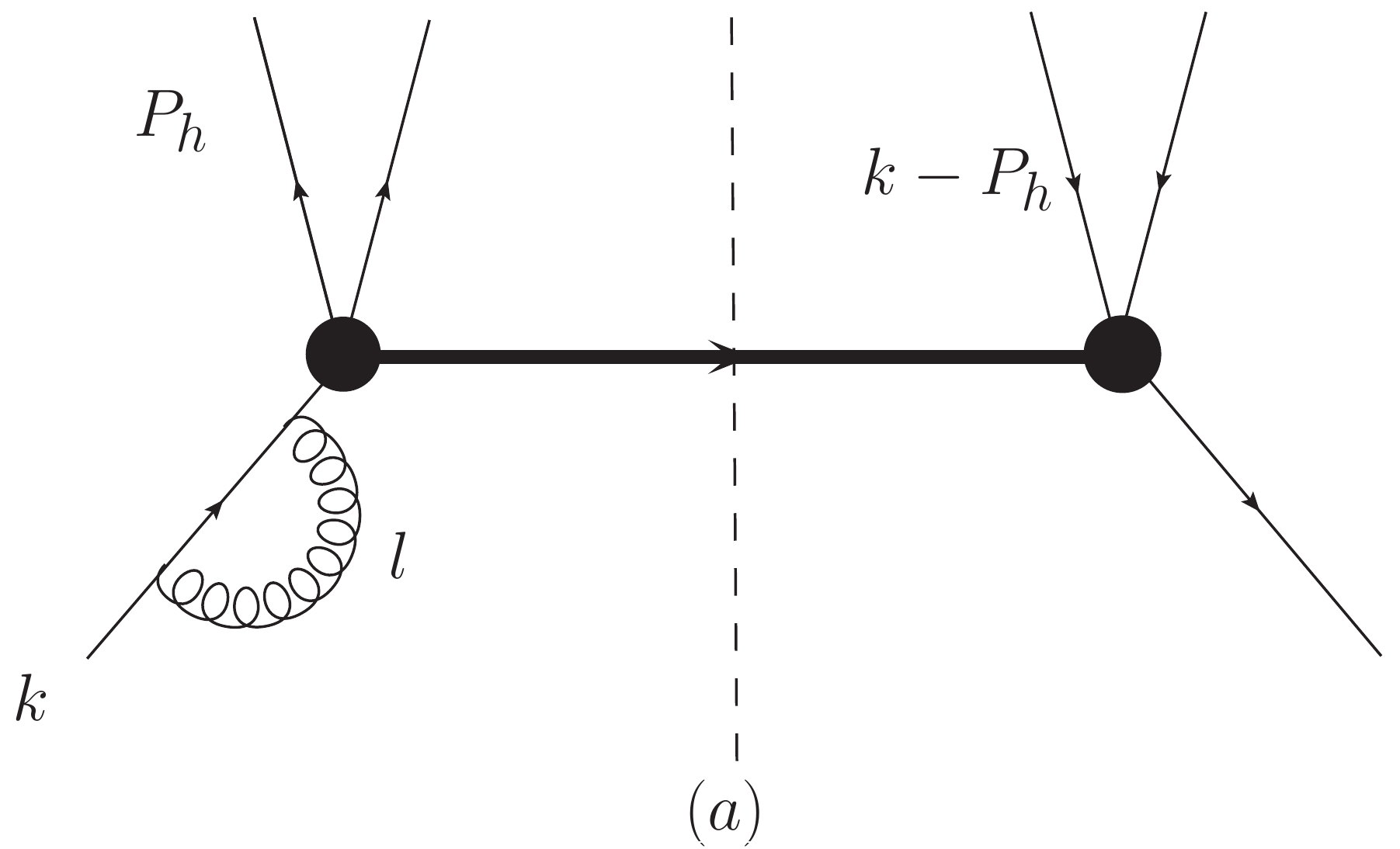}
\includegraphics[height=3cm,width=4cm]{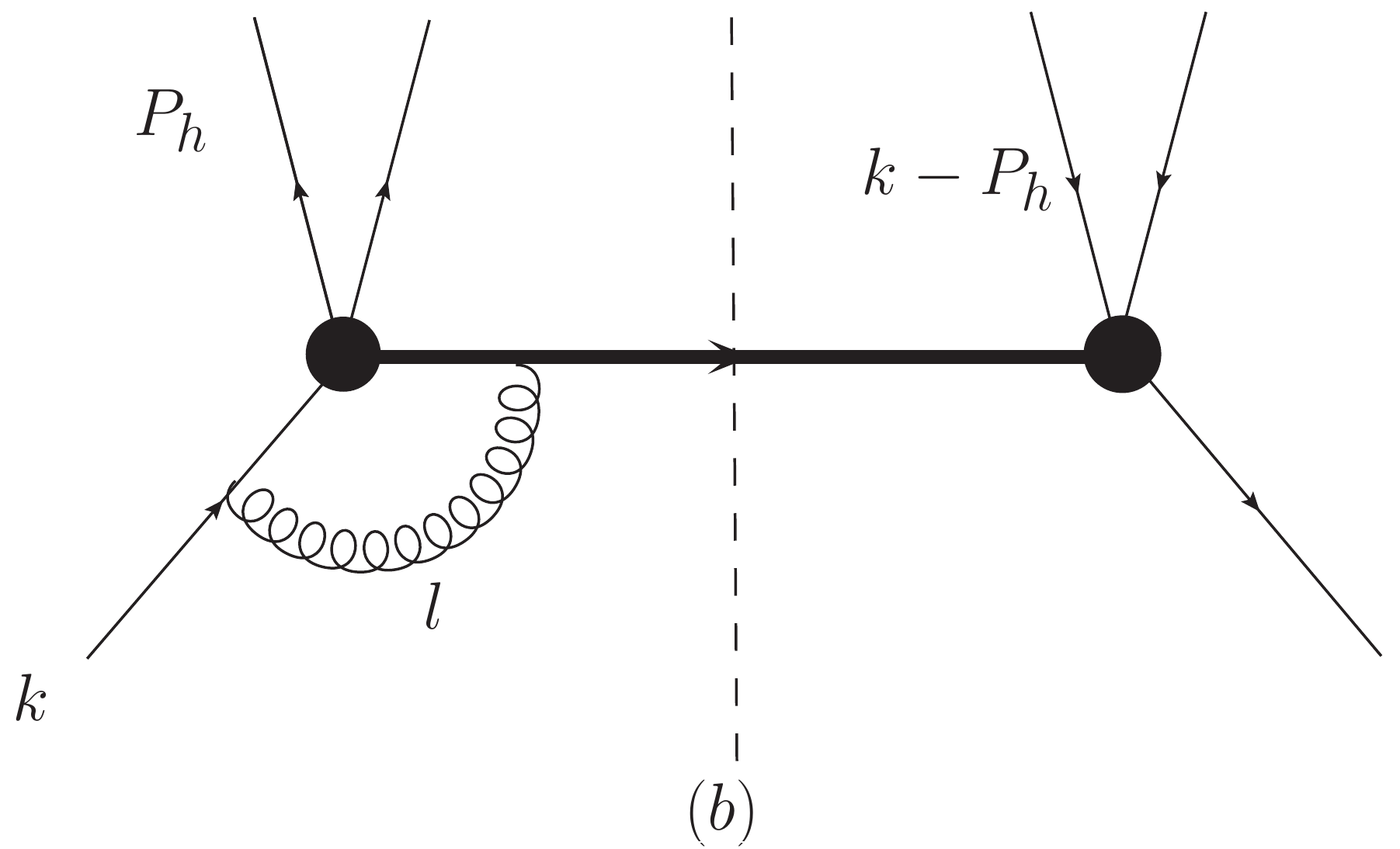}
\includegraphics[height=3cm,width=4cm]{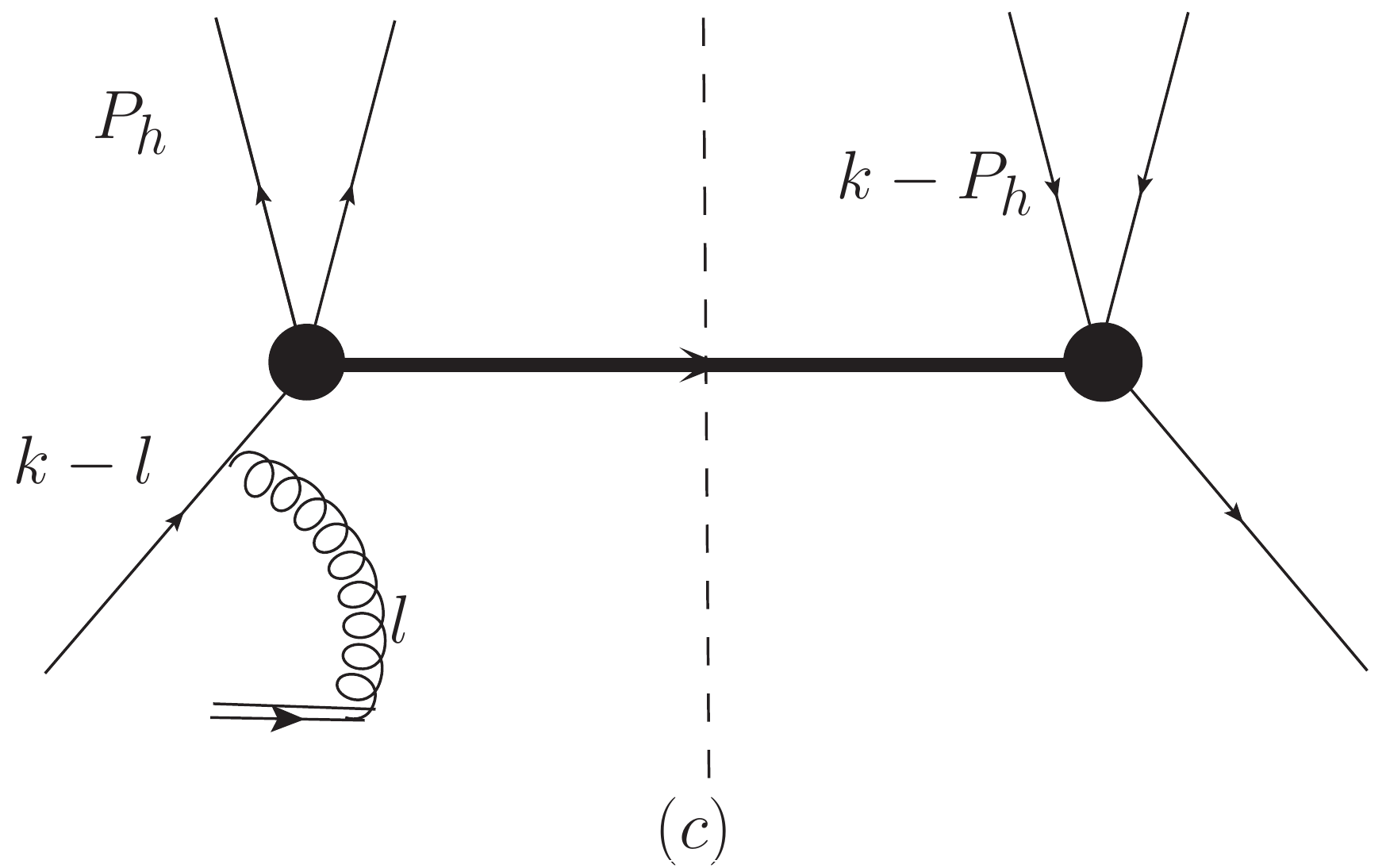}
\includegraphics[height=3cm,width=4cm]{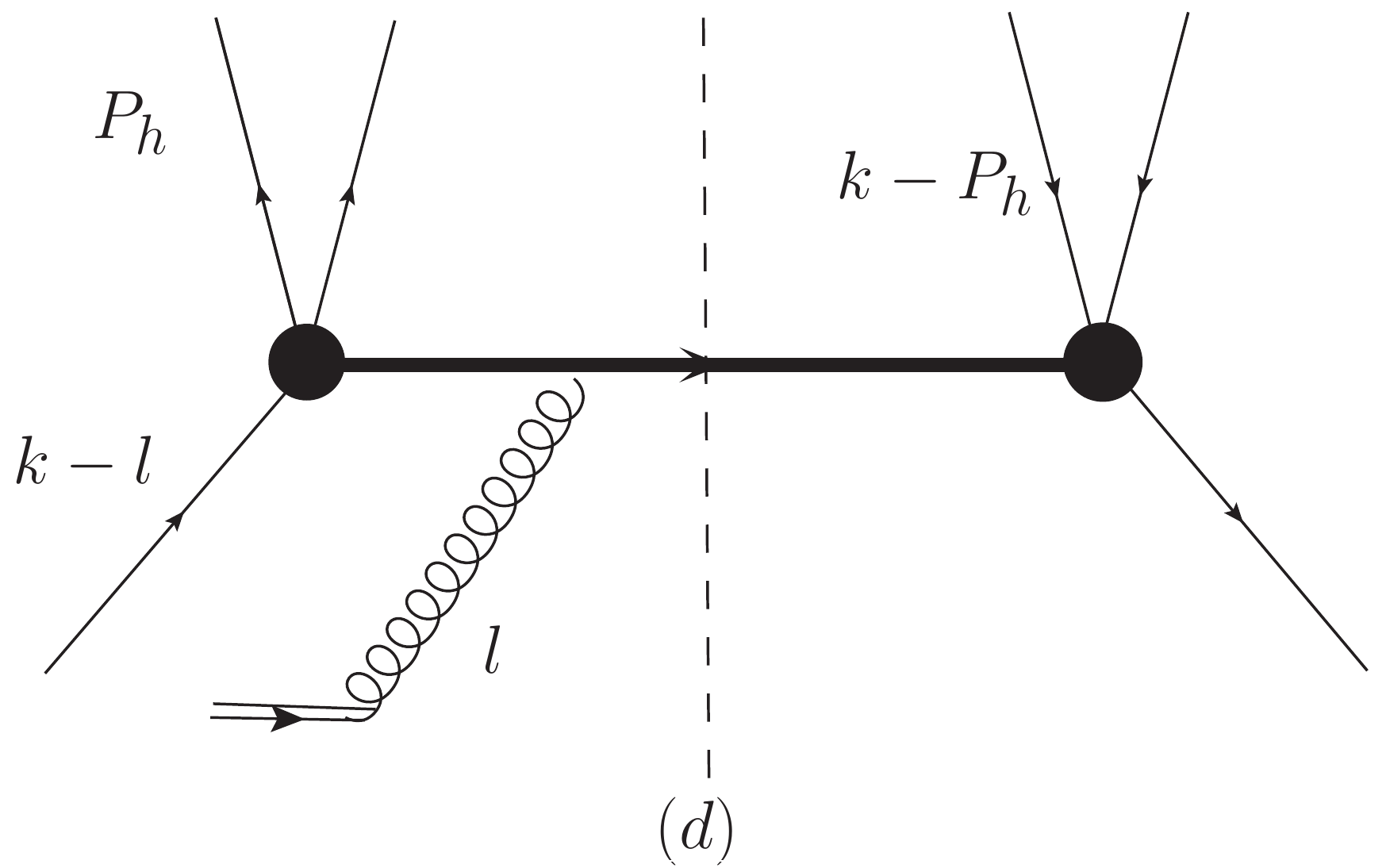}
+h.c.
\caption{ One loop order corrections to the fragmentation function of a quark into a meson pair in the spectator model. 
Where h.c. represents the hermitian conjugations of these diagrams.}
\label{fig:1}
\end{figure}
We can model the correlator at one loop level provided with Fig.\ref{fig:1} as:
\begin{eqnarray} \label{eq:15} 
\begin{aligned}
\Delta_a^q(z,k_T^2,\cos\theta,M_h^2,\phi_R) &= i\frac{C_F \alpha_s}{32\pi^2 (1-z) P_h^-} 
\cdot \frac{ |\vec{R}|}{M_h} \cdot \frac{(\s{k}+m)}{(k^2-m^2)^3}\left( F^{s*}e^{-\frac{k^2}{\Lambda_s^2}}
+F^{p*}e^{-\frac{k^2}{\Lambda_p^2}} \s{R} \right) (\s{k}-\s{P}_h+M_s) 
\\
&\times \left( F^{s}e^{-\frac{k^2}{\Lambda_s^2}}+F^{p}e^{-\frac{k^2}{\Lambda_p^2}} \s{R} \right) (\s{k}+m)
\int \frac{d^4 \ell}{(2\pi)^4} \frac{\gamma^\mu (\s{k}-\s{\ell}+m) \gamma_\mu (\s{k}+m)}{((k-\ell)^2-m^2+i\varepsilon)(\ell^2+i\varepsilon)} ,
\end{aligned}
\end{eqnarray}
\begin{eqnarray} \label{eq:16}
\begin{aligned} 
\Delta_b^q(z,k_T^2,\cos\theta,M_h^2,\phi_R) &= i\frac{C_F \alpha_s}{32\pi^2 (1-z) P_h^-} 
\cdot \frac{ |\vec{R}|}{M_h} \cdot \frac{(\s{k}+m)}{(k^2-m^2)^2}
\left( F^{s*}e^{-\frac{k^2}{\Lambda_s^2}}+F^{p*}e^{-\frac{k^2}{\Lambda_p^2}} \s{R} \right) (\s{k}-\s{P}_h+M_s) 
\\
&\times \int \frac{d^4 \ell}{(2\pi)^4} \frac{\gamma^\mu (\s{k}-\s{P}_h-\s{\ell}+M_s) 
\Big( F^{s}e^{-\frac{k^2}{\Lambda_s^2}}+F^{p}e^{-\frac{k^2}{\Lambda_p^2}} \s{R} \Big)  
(\s{k}-\s{\ell}+m) \gamma_\mu (\s{k}+m)}{((k-P_h-\ell)^2-M_s^2+i\varepsilon)((k-\ell)^2-m^2+i\varepsilon)(\ell^2+i\varepsilon)} ,
\end{aligned}
\end{eqnarray}
\begin{eqnarray} \label{eq:17} 
\begin{aligned}
\Delta_c^q(z,k_T^2,\cos\theta,M_h^2,\phi_R) &= i\frac{C_F \alpha_s}{32\pi^2 (1-z) P_h^-} 
\cdot \frac{ |\vec{R}|}{M_h} \cdot \frac{(\s{k}+m)}{(k^2-m^2)^2}\left( F^{s*}e^{-\frac{k^2}{\Lambda_s^2}}+F^{p*}e^{-\frac{k^2}{\Lambda_p^2}} \s{R} \right) (\s{k}-\s{P}_h+M_s) 
\\
&\times \left( F^{s}e^{-\frac{k^2}{\Lambda_s^2}}+F^{p}e^{-\frac{k^2}{\Lambda_p^2}} \s{R} \right) 
\int \frac{d^4 \ell}{(2\pi)^4} \frac{(\s{k}+m) \gamma^- (\s{k}-\s{\ell}+m) }{((k-\ell)^2-m^2+i\varepsilon)(-\ell^- \pm i\varepsilon)(\ell^2+i\varepsilon)} ,
\end{aligned}
\end{eqnarray}
\begin{eqnarray} \label{eq:18} 
\begin{aligned}
\Delta_d^q(z,k_T^2,\cos\theta,M_h^2,\phi_R) &= i\frac{C_F \alpha_s}{32\pi^2 (1-z) P_h^-} 
\cdot \frac{ |\vec{R}|}{M_h} \cdot \frac{(\s{k}+m)}{k^2-m^2}\left( F^{s*}e^{-\frac{k^2}{\Lambda_s^2}}+F^{p*}e^{-\frac{k^2}{\Lambda_p^2}} \s{R} \right) (\s{k}-\s{P}_h+M_s) 
\\
&\times \int \frac{d^4 \ell}{(2\pi)^4} \frac{\gamma^- (\s{k}-\s{P}_h-\s{\ell}+M_s) 
\Big( F^{s}e^{-\frac{k^2}{\Lambda_s^2}}+F^{p}e^{-\frac{k^2}{\Lambda_p^2}} \s{R} \Big)  
(\s{k}-\s{\ell}+m) }{((k-P_h-\ell)^2-M_s^2+i\varepsilon)((k-\ell)^2-m^2+i\varepsilon)(-\ell^- \pm i\varepsilon)(\ell^2+i\varepsilon)} .
\end{aligned}
\end{eqnarray}
In Eq.(\ref{eq:15}-\ref{eq:18}) we have applied the Feynman rule $1/(-\ell^- \pm i\varepsilon)$ for the eikonal propagator, 
as well as that for the vertex between the eikonal line and the gluon.
Still in Eq.(\ref{eq:15}-\ref{eq:18}), in principle, the Gaussian form factors should depend on the loop momentum $\ell$. 
Following the choice made in Ref.\cite{Bacchetta:2007wc}, we abandon this dependence and simply use $k^2$ in those form factors
instead of $(k-\ell)^2$ in order to simplify the integration. This choice could give reasonable final results since the form factor is introduced to cut off the divergence. 
The same choice has also been adopted in Refs. \cite{Bacchetta:2002tk,Bacchetta:2003xn,Amrath:2005gv}.

In general, there are two sources of $G_{1,OT}^\perp$ in each diagram at one loop level.
One is the real part of the loop integral over $\ell$, coupled with the imaginary part of $F^* F^p$. 
The other is the imaginary part of the loop integral, combined with the real part of $F^* F^p$. 
The real part of the integral is presented as the usual loop integral adopting the Feynman parameterization. 
While for the imaginary part of the integral, we apply the Cutkosky cutting rules
\begin{eqnarray} \label{eq:19}
\begin{aligned}
\frac{1}{\ell^2+i\varepsilon} \to -2\pi i \delta(\ell^2) \qquad \qquad \frac{1}{(k-\ell)^2+i\varepsilon} \to -2\pi i \delta((k-\ell)^2) .
\end{aligned}
\end{eqnarray}

Employing the above conventions, we reach the final result of $G_{1,OT}^\perp$
\begin{eqnarray} \label{eq:20} 
\begin{aligned}
G_{1,OT}^{\perp a}&=0
\\
G_{1,OT}^{\perp b}&=-\frac{1}{8\pi^3}\left[ \frac{C_F \alpha_s M_h |\vec{R}|}{(1-z)} 
\cdot \Re[F^{s*} F^p] e^{-\frac{2k^2}{\Lambda_{sp}^2}} \right]\frac{1}{(k^2-m^2)^2} C_b 
\\
&+ \frac{1}{16\pi^3}\left[ \frac{C_F \alpha_s M_h M_s |\vec{R}|}{(1-z)} 
\cdot \Im[F^{s*} F^p] e^{-\frac{2k^2}{\Lambda_{sp}^2}} \right]\frac{1}{(k^2-m^2)^2} k^2 \cdot C_2(k^2,M_h^2,2k^2+2M_h^2-M_s^2,0,0,M_s)
\\
G_{1,OT}^{\perp c}&=0
\\
G_{1,OT}^{\perp d}&=\frac{1}{2\pi^3}\left[ \frac{C_F \alpha_s M_h M_s |\vec{R}|}{(1-z)} 
\cdot \Re[F^{s*} F^p] e^{-\frac{2k^2}{\Lambda_{sp}^2}} \right] \frac{1}{k^2-m^2} (-\mathcal{D} P_h^-)
\end{aligned}
\end{eqnarray}
with
\begin{eqnarray} \label{eq:21}
\begin{aligned}
C_b &= (4k^2M_s-2mk^2-4m^3-4m^2M_s)\mathcal{A}+(2k^2M_s-mk^2+4M_h^2M_s-3mM_h^2-2m^3+2m^2 M_s
\\
& +3mM_s^2-4M_s^3)\mathcal{B}+(2mk^2+2mM_h^2-2mM_s^2)\mathcal{B}_0+4mM_h^2 \mathcal{D}_0+(3mk^2-3m^3)I_2+8m\mathcal{E}_0 ,
\end{aligned}
\end{eqnarray}
where $C_2$ is a three point one loop tensor integration defined as
\begin{eqnarray} \label{eq:22}
\begin{aligned}
&C_2(p_1^2,p_2^2,p_3^2;m_1^2,m_2^2,m_3^2)=\bigg( -2p_1^2 B_0(p_1^2,m_1^2,m_2^2)+(p_1^2+p_2^2-p_3^2)B_0(p_2^2,m_2^2,m_3^2)+(p_1^2-p_2^2+p_3^2)B_0(p_3^2,m_1^2,m_3^2)
\\
&+[m_1^2(p_1^2+p_2^2-p_3^2)+m_2^2(p_1^2-p_2^2+p_3^2)+p_1^2(-2m_3^2-p_1^2+p_2^2+p_3^2)]C_0(p_1^2,p_2^2,p_3^2;m_1^2,m_2^2,m_3^2) \bigg)
\\
&\times \frac{1}{p_1^4-2(p_2^2+p_3^2)p_1^2+(p_2^2-p_3^2)^2}
\end{aligned}
\end{eqnarray}
where the general defination of 2-point one-loop tensor integration $B_0$ is given by \cite{Ellis:2007qk} 
\begin{eqnarray} \label{eq:23}
\begin{aligned}
B_0(p_1^2;m_1^2,m_2^2)=\frac{1}{i\pi^2} \int d^4 l \frac{1}{(l^2-m_1^2+i\varepsilon)((l+q_1)^2-m_2^2+i\varepsilon)}
\end{aligned}
\end{eqnarray}
and 3-point one-loop tensor integration $C_0$ is denoted as \cite{Ellis:2007qk} 
\begin{eqnarray} \label{eq:24}
\begin{aligned}
C_0(p_1^2,p_2^2,p_3^2;m_1^2,m_2^2,m_3^2)=\frac{1}{i\pi^2} \int d^4 \ell \frac{1}{(\ell^2-m_1^2+i\varepsilon)((\ell+q_1)^2-m_2^2+i\varepsilon)((\ell+q_2)^2-m_3^2+i\varepsilon)} 
\end{aligned}
\end{eqnarray}
with $q_n \equiv \sum_{i=1}^n p_i$ and $q_0=0$.
The coefficients $\mathcal{A}$ and $\mathcal{B}$ denote the following functions
\begin{eqnarray} \label{eq:25}
\begin{aligned}
\mathcal{A} & =\frac{I_1}{\lambda(M_h,M_s)}\left[ 2k^2(k^2-M_s^2-M_h^2)\frac{I_2}{\pi}+(k^2+M_h^2-M_s^2) \right] \\
\mathcal{B} & =-\frac{2k^2}{\lambda(M_h^2,M_s^2)}I_1\left[ 1+\frac{k^2+M_s^2-M_h^2}{\pi}I_2 \right]
\end{aligned}
\end{eqnarray}
which originate from the decomposition of the following integral \cite{Lu:2015wja}
\begin{eqnarray} \label{eq:26}
\begin{aligned}
\int d^4 \ell \frac{\ell^\mu \delta(\ell^2) \delta[(k-\ell)^2-m^2]}{(k-P_h-\ell)^2-M_s^2}=\mathcal{A}k^\mu + \mathcal{B}P_h^\mu .
\end{aligned}
\end{eqnarray}
The functions $I_i$ represent the results of the following integrals
\begin{eqnarray} \label{eq:27}
\begin{aligned}
I_1&=\int d^4 \ell \delta(\ell^2) \delta[(k-\ell)^2-m^2]=\frac{\pi}{2k^2}(k^2-m^2) \\
I_2&=\int d^4 \ell \frac{\delta(\ell^2)\delta[(k-\ell)^2-m^2]}{(k-\ell-P_h)^2-M_s^2}
=\frac{\pi}{2\sqrt{\lambda(M_h,M_s)}}\ln\left( 1-\frac{2\sqrt{\lambda(M_h,M_s)}}{k^2-M_h^2+M_s^2+\sqrt{\lambda(M_h,M_s)}} \right)
\\
I_3&=\int d^4 \ell \frac{\delta(\ell^2)\delta((k-\ell)^2-m^2)}{-\ell^-+i\varepsilon}
\\
I_4&=\int d^4 \ell \frac{\delta(\ell^2)\delta((k-\ell)^2-m^2)}{(-\ell^-+i\varepsilon)((k-P_h-\ell)^2-M_s^2)}
\end{aligned}
\end{eqnarray}
with $\lambda(M_h,M_s)=[k^2-(M_h+M_s)^2][k^2-(M_h-M_s)^2]$. The function $\mathcal{D}$ satisfies the following relation
\begin{eqnarray} \label{eq:28}
\begin{aligned}
P_h^- \mathcal{D} &= -\frac{I_{34}}{2zk_T^2}-\frac{1}{2zk_T^2}\left( (1-2z)k^2+M_h^2-M_s^2 \right)I_2 ,
\end{aligned}
\end{eqnarray}
where $I_{34}$ is the linear combination of $I_3$ and $I_4$
\begin{eqnarray} \label{eq:29}
\begin{aligned}
I_{34}=k^- (I_3+(1-z)(k^2-m^2)I_4) = \pi\ln\frac{\sqrt{k^2}(1-z)}{M_s} .
\end{aligned}
\end{eqnarray}
In addition, the function $\mathcal{B}_0$ and $\mathcal{D}_0$ come from the decomposition
\begin{eqnarray} \label{eq:30}
\begin{aligned}
\int d^4 \ell \frac{\ell^\mu \ell^\nu \delta(\ell^2) \delta((k-\ell)^2-m^2)}{(k-p-\ell)^2-M_s^2} 
&= k^\mu k^\nu \mathcal{A}_0+k^\mu p^\nu \mathcal{B}_0+p^\mu k^\nu \mathcal{C}_0+p^\mu p^\nu \mathcal{D}_0+g^{\mu\nu}\mathcal{E}_0 ,
\end{aligned}
\end{eqnarray}
where
\begin{eqnarray} \label{eq:31}
\begin{aligned}
\mathcal{B}_0&=\frac{1}{2}\frac{(k^2-m^2)(\mathcal{A}k^2+3\mathcal{B}k^2+\mathcal{A}M_h^2
-\mathcal{B}M_h^2-\mathcal{A}M_s^2-3\mathcal{B}M_s^2)}{k^4-2k^2M_h^2-2k^2M_s^2+M_h^4-2M_h^2M_s^2+M_s^4}
\\
\mathcal{D}_0&=-\frac{(k^2-m^2)(\mathcal{A}k^2+2\mathcal{B}k^2-\mathcal{B}M_h^2+\mathcal{B}M_s^2)}{k^4-2k^2M_h^2-2k^2M_s^2+M_h^4-2M_h^2M_s^2+M_s^4} 
\\
\mathcal{E}_0&=-\frac{1}{4}(k^2-m^2)(\mathcal{A}+\mathcal{B})
\end{aligned}
\end{eqnarray}

\section{The $\sin(\phi_h-\phi_R)$ asymmetry of dihadron production in SIDIS}
\label{IV}

\begin{figure}[htp]
\centering
\includegraphics[height=6.6cm,width=10.4cm]{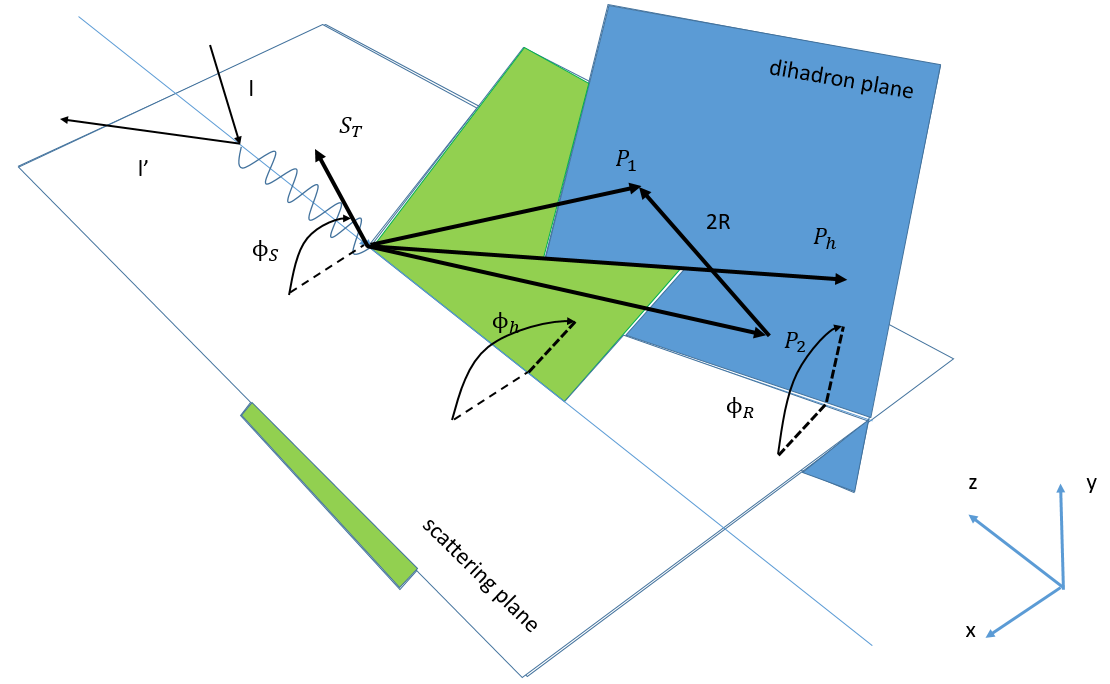}
\caption{Angle definitions involved in the measurement of the single longitudinal spin asymmetry in SIDIS production of two hadrons.}
\label{fig:2}
\end{figure}
The reference system in Fig.\ref{fig:2} is our starting point in which the virtual photon transverse momentum is set to $\vec{q}_T=0$. But in our work, the virtual photon has a nonvanishing transverse momentum $\vec{q}_T$. We align the $\hat{z}$ axis opposite to the direction of the virtual photon
for the transverse momentum dependence case and, the outgoing hadron has a nonvanishing transverse momentum defined as $\vec{P}_{h\perp} = -z \vec{q}_T$. 
Then we consider the dihadron SIDIS production
\begin{eqnarray} \label{eq:32}
\begin{aligned}
\mu(\ell)+p^\rightarrow(P) \to \mu(\ell')+h^+(P_1)+h^-(P_2)+X ,
\end{aligned}
\end{eqnarray}
where a unpolarized muon with momentum $\ell$ scatters off a longitudinally polarized target nucleon with mass $M$, 
polarization $S$ and momentum $P$, via the exchange of a virtual photon with momentum $q=\ell-\ell'$. 
Inside the target, the photon hits the active quark with momentum $p$ and the final state quark with momentum $k=p+q$ 
then fragments into two leading unpolarized hadrons with mass $M_1, M_2$, and momenta $P_1, P_2$. 
To present the differential cross section with respect to dihadron-dependent structure function, we define the following kinematic invariants:
\begin{eqnarray} \label{eq:33}
\begin{aligned}
x   &=\frac{k^+}{P^+} \qquad  y=\frac{P \cdot q}{P \cdot \ell} \qquad z=\frac{P_h^-}{k^-}=z_1+z_2 \\
z_i &=\frac{P_i^-}{k^-} \qquad  Q^2=-q^2  \qquad s=(P+\ell)^2 \\
P_h &=P_1+P_2 \qquad  R=\frac{P_1-P_2}{2} \qquad  M_h^2=P_h^2  .
\end{aligned}
\end{eqnarray}
The light-cone fraction of target momentum carried by the initial quark is denoted by $x$, $z_i$ 
expresses the light-cone fraction of hadron $h_i$ in terms of the fragmented quark.
The light-cone fraction of fragmenting quark momentum carried by the final hadron pair is defined by $z$. 
Moreover, the invariant mass, the total momentum and the relative momentum of the hadron pair are denoted by $M_h$, $P_h$ and $R$, respectively.

We will consider the SIDIS process with unpolarized muons off longitudinally polarized nucleon target. 
Using TMD factorization approach and denoting $A(y) = 1-y+\frac{y^2}{2}$, the differential cross section for this process reads \cite{Radici:2001na}
\begin{eqnarray} \label{eq:34}
\begin{aligned}
\displaystyle \frac{d^9 \sigma_{UU}}{dx dy dz d\phi_S d\phi_h d\phi_R d\cos\theta d\vec{P}_{h\perp}^2 dM_h^2} 
= \frac{\alpha^2}{2\pi s x y^2} A(y) \sum_q e_q^2 \mathcal{I}[f_1^q D_{1,OO}^q]
\end{aligned}
\end{eqnarray}
and
\begin{eqnarray} \label{eq:35} 
\begin{aligned}
\displaystyle \frac{d^9 \sigma_{UL}}{dx dy dz d\phi_S d\phi_h d\phi_R d\cos\theta d\vec{P}_{h\perp}^2 dM_h^2} 
&=\frac{\alpha^2}{2\pi s x y^2} A(y) \sum_q e_q^2 \bigg\{ \sin\theta \sin(\phi_h-\phi_R) 
\mathcal{I}\left[ \frac{\vec{k}_T \cdot \hat{P}_{h\perp}}{M_h} g_{1L}^q \left( \frac{|\vec{R}|}{M_h} G_{1,OT}^\perp \right) \right] 
\\
&+ \sin 2\theta \sin(\phi_h-\phi_R) \mathcal{I}\left[ \frac{\vec{k}_T 
\cdot \hat{P}_{h\perp}}{M_h} g_{1L}^q \left( \frac{|\vec{R}|}{2M_h} G_{1,LT}^\perp \right) \right] \bigg\} , 
\end{aligned}
\end{eqnarray}
where $\phi_R$ and $\phi_S$ are the azimuthal angles of $\vec{R}_T$ and $\vec{S}_T$ with respect to the lepton scattering plane. 
$\hat{P}_{h\perp}$ satisfies $\hat{P}_{h\perp} = \vec{P}_{h\perp}/|\vec{P}_{h\perp}|$.
For convenience, we have indicated the unpolarized or longitudinally polarized states of the beam or the target with the labels $U$ and $L$, respectively. 
The structure functions occuring in Eqs.(\ref{eq:34}-\ref{eq:35}) are written as weighted convolutions of the form
\begin{eqnarray} \label{eq:36}
\begin{aligned}
\mathcal{I}[f] = \int d^2 \vec{p}_T d^2 \vec{k}_T \delta\left( \vec{p}_T-\vec{k}_T-\frac{\vec{P}_{h\perp}}{z} \right)[f] .
\end{aligned}
\end{eqnarray}
In Eq.(\ref{eq:34}), $f_1^q$ and $D_{1,OO}^q$ are the unpolarized PDF and unpolarized DiFF with flavor $q$. 
In Eq.(\ref{eq:35}), $g_{1L}^q$ is a twist-2 distribution function coupled with the T-odd DiFF $G_{1,OT}^{\perp}$ and $G_{1,LT}^{\perp}$ respectively. 
In principle, both DiFFs contribute to the $\sin(\phi_h-\phi_R)$ azimuthal asymmetry in SIDIS. 
However, to access the asymmetry one must do the integration with respect to $\theta$ 
and we have $\int_{-1}^{1} \sin2\theta d\cos\theta = 0$ for the latter contribution. 
Thus the contribution including $G_{1,LT}^\perp$ vanishes and the $\sin(\phi_h-\phi_R)$ asymmetry of the considered process can be expressed as
\begin{eqnarray} \label{eq:37}
\begin{aligned}
A_{UL}^{\sin(\phi_h-\phi_R)}&=&\displaystyle \frac{\pi}{4} \frac{\sum_q e_q^2 
\mathcal{I}\left[ \frac{\vec{k}_T \cdot \hat{P}_{h\perp}}{M_h} g_{1L}^q 
\left( \frac{|\vec{R}|}{M_h} G_{1,OT}^\perp \right) \right]}{\sum_q e_q^2 \int \mathcal{I}[f_1^q D_{1,OO}^q]} .
\end{aligned}
\end{eqnarray}

\section{Numerical results}
\label{V}

In order to fix the parameters of the spectator model, the authors of Ref. \cite{Bacchetta:2006un} 
compare it with the output of the PYTHIA event generator \cite{Sjostrand:2000wi} adopted for HERMES. The values of the parameters obtained by the fit are:
\begin{alignat*}{3}
\alpha_s &=2.60\ \text{GeV} &\qquad& \beta_s=-0.751 &\qquad& \gamma_s=-0.193 \nonumber \\
\alpha_p &=7,07\ \text{GeV} &\qquad& \beta_p=-0.038 &\qquad& \gamma_p=-0.085 \\
f_s &=1197\ \text{GeV}^{-1} &\qquad& f_\rho=93.5 &\qquad& f_\omega=0.63 \nonumber\\
f_\omega' &=75.2 &\qquad& M_s=2.97M_h &\qquad& m=0.0\ \text{GeV} , \nonumber
\end{alignat*}
where we have adopted the same choice as in Ref. \cite{Bacchetta:2006un} for the quark mass $m$ fixed to be zero GeV. 
It's desired to mention that these model parameters are obtained by comparing the theoretical model 
with the PITHIA event generator adopted for the HERMES kinematics.
In the following we will make predictions in COMPASS and EIC kinematics, thus there are uncertainties with respect to the model parameters.
In this paper, we make a rough consideration by ignoring such uncertainties.
Furthermore, we make a preliminary estimate for choosing the strong coupling $\alpha_s \approx 0.3$.

\begin{figure}[htp]
\centering
\includegraphics[height=6.0cm,width=7.4cm]{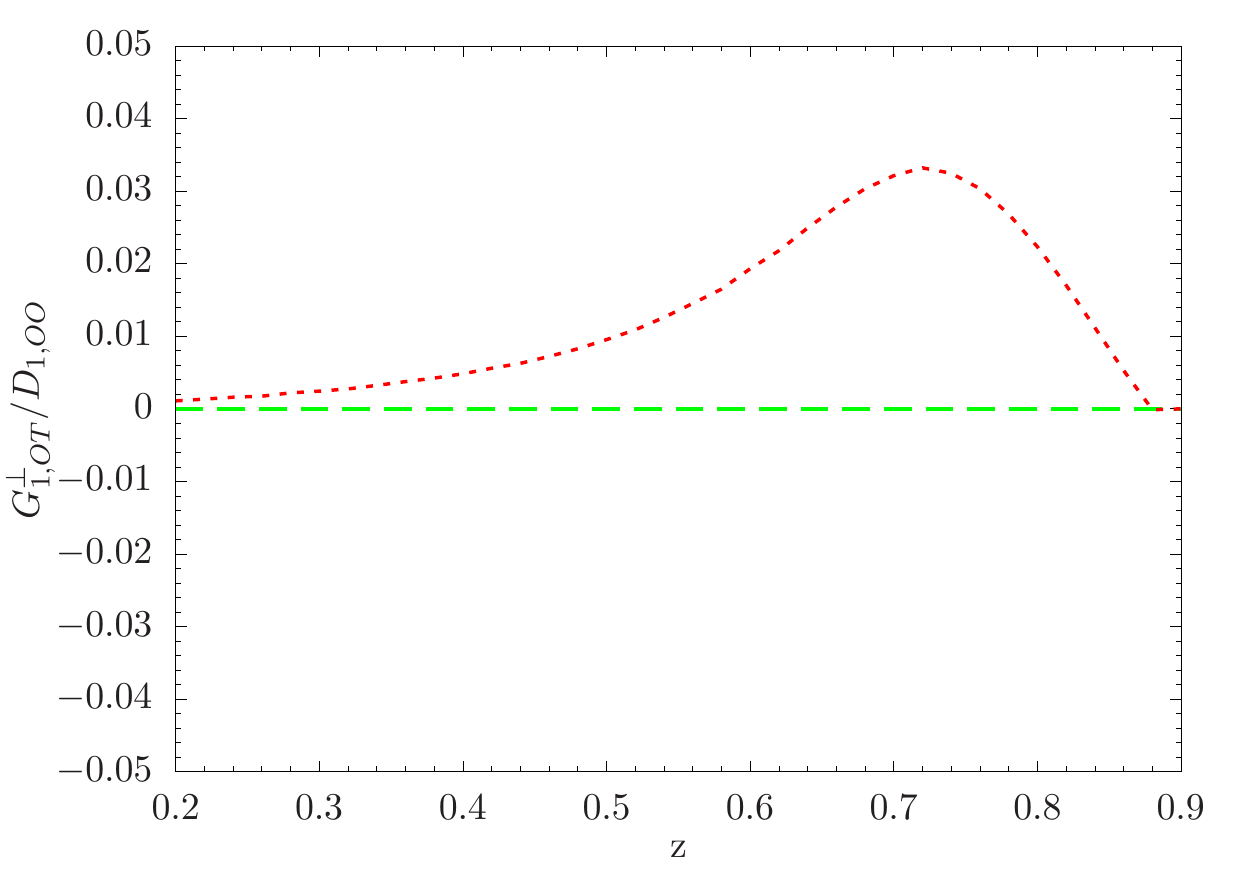}
\includegraphics[height=6.0cm,width=7.4cm]{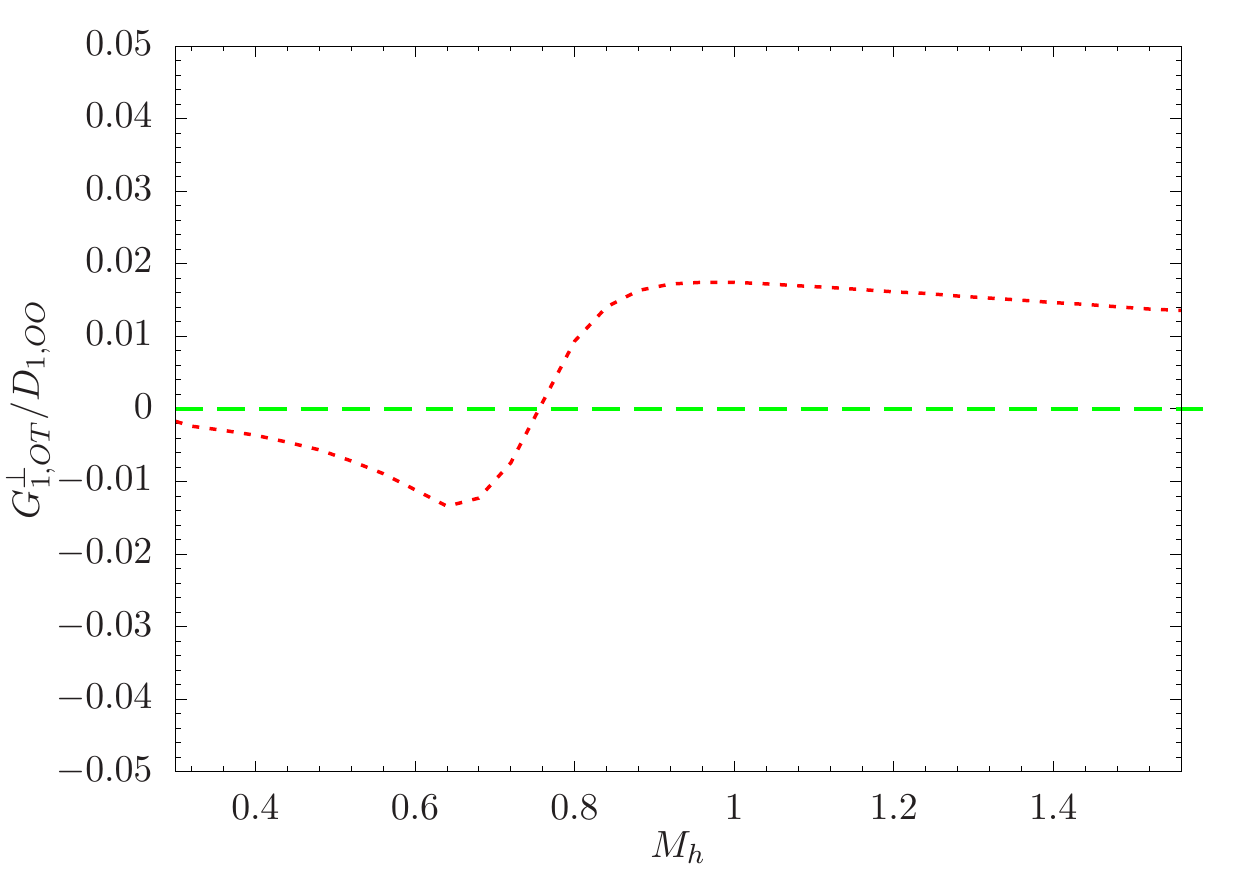}
\caption{\normalsize 
The DiFF $G^\perp_{1,OT}$ as functions of $z$ (left panel) and $M_h$ (right panel) in the spectator model, normalized by the unpolarized DiFF $D_{1,OO}$.}
\label{fig:3}
\end{figure}
We map out the ratio between $G^\perp_{1,OT}$ and $D_{1,OO}$ as a function of $z$ or $M_h$, 
integrated over the region $0.3\ \text{GeV}<M_h<1.6\ \text{GeV}$ or $0.2<z<0.9$ 
in the left panel and right panel  of Fig.\ref{fig:3} respectively. 
Comparing with the unpolarized DiFF $D_{1,OO}$, the $G^\perp_{1,OT}$ 
is two order of magnitude smaller and we can find a break point located nearly in $M_h=0.8$.

Then we present the numerical results of the $\sin(\phi_h-\phi_R)$ azimuthal asymmetry in the SIDIS process 
with unpolarized muons off longitudinally polarized nucleon target. 
According to isospin symmetry we have the conclusion that the fragmentation correlators 
for processes $u \to \pi^+ \pi^- X$, $\bar{d} \to \pi^+ \pi^- X$, $d \to \pi^- \pi^+ X$ and $\bar{u} \to \pi^- \pi^+ X$ are the same.
Thus by transforming the sign of $\vec{R}$, equivalently changing $\theta \to \pi-\theta$ 
and $\phi \to \phi+\pi$, the DiFF $G_{1,OT}^\perp$ which depends linearly on $R$ 
coming from $d \to \pi^- \pi^+ X$ and $\bar{u} \to \pi^- \pi^+ X$ processes has an additional minus sign comparing to the $u \to \pi^+ \pi^- X$ process.
When expanding the flavor sum in the numerator of Eq.(\ref{eq:37}), 
we apply the isospin symmetry to the DiFF $G_1^\perp$. Furthermore, 
in principle sea quark distributions can be generated via perturbative QCD evolution and they are zero at the model scale. 
In this paper, we make a rough consideration by ignoring QCD evolution, which leads to zero antiquark PDFs $f_1$ and $g_{1L}$.
The expressions of the $x$-dependent, $z$-dependent and $M_h$-dependent $\sin(\phi_h-\phi_R)$ asymmetry can be adopted from Eq.(\ref{eq:37}) as follows
\begin{eqnarray} \label{eq:38}
\begin{aligned}
&\displaystyle A_{UL}^{\sin(\phi_h-\phi_R)}(x) = \frac{\int dz 2M_h dM_h  
d\cos\theta d^2 \vec{P}_{h\perp} d^2 \vec{p}_T d^2 \vec{k}_T \delta\left( \vec{p}_T-\vec{k}_T-\frac{\vec{P}_{h\perp}}{z}  
\right) (4g_{1L}^u(\vec{p}_T^2)-g_{1L}^d(\vec{p}_T^2)) G_{1,OT}^\perp(\vec{k}_T^2) \sin\theta \frac{\vec{k}_T 
\cdot \vec{P}_{h\perp} |\vec{R}|}{2M_h^2 |\vec{P}_{h\perp}|}}{\int dz 2M_h dM_h  d\cos\theta d^2 \vec{P}_{h\perp} d^2 
\vec{p}_T d^2 \vec{k}_T \delta\left( \vec{p}_T-\vec{k}_T-\frac{\vec{P}_{h\perp}}{z}  \right) (4f_{1}^u(\vec{p}_T^2)+f_{1}^d(\vec{p}_T^2)) D_{1,OO}(\vec{k}_T^2)}
\\
&\displaystyle A_{UL}^{\sin(\phi_h-\phi_R)}(z) = \frac{\int dx 2M_h dM_h  
d\cos\theta d^2 \vec{P}_{h\perp} d^2 \vec{p}_T d^2 \vec{k}_T \delta\left( \vec{p}_T-\vec{k}_T-\frac{\vec{P}_{h\perp}}{z}  
\right) (4g_{1L}^u(\vec{p}_T^2)-g_{1L}^d(\vec{p}_T^2)) G_{1,OT}^\perp(\vec{k}_T^2) \sin\theta \frac{\vec{k}_T 
\cdot \vec{P}_{h\perp} |\vec{R}|}{2M_h^2 |\vec{P}_{h\perp}|}}{\int dx 2M_h dM_h  d\cos\theta d^2 \vec{P}_{h\perp} d^2 
\vec{p}_T d^2 \vec{k}_T \delta\left( \vec{p}_T-\vec{k}_T-\frac{\vec{P}_{h\perp}}{z}  \right) (4f_{1}^u(\vec{p}_T^2)+f_{1}^d(\vec{p}_T^2)) D_{1,OO}(\vec{k}_T^2)}
\\
&\displaystyle A_{UL}^{\sin(\phi_h-\phi_R)}(M_h) = \frac{\int dx dz 
d\cos\theta d^2 \vec{P}_{h\perp} d^2 \vec{p}_T d^2 \vec{k}_T \delta\left( \vec{p}_T-\vec{k}_T-\frac{\vec{P}_{h\perp}}{z}  
\right) (4g_{1L}^u(\vec{p}_T^2)-g_{1L}^d(\vec{p}_T^2)) G_{1,OT}^\perp(\vec{k}_T^2) \sin\theta \frac{\vec{k}_T 
\cdot \vec{P}_{h\perp} |\vec{R}|}{2M_h^2 |\vec{P}_{h\perp}|}}{\int dx dz d\cos\theta d^2 \vec{P}_{h\perp} d^2 \vec{p}_T d^2 
\vec{k}_T \delta\left( \vec{p}_T-\vec{k}_T-\frac{\vec{P}_{h\perp}}{z}  \right) (4f_{1}^u(\vec{p}_T^2)+f_{1}^d(\vec{p}_T^2)) D_{1,OO}(\vec{k}_T^2)} ,
\end{aligned}
\end{eqnarray}
where the TMD DiFF $D_{1,OO}$ has been worked out and listed as
\begin{eqnarray} \label{eq:39}
\begin{aligned}
D_{1,OO}(z,\vec{k}_T^2,M_h)=&\frac{4\pi|\vec{R}|}{256\pi^3M_h z (1-z)(k^2-m^2)^2}\bigg\{ 4|F^s|^2 e^{-\frac{2k^2}{\Lambda_s^2}} (zk^2-M_h^2-m^2z+m^2+2mM_s+M_s^2)
\\
&-4|F^p|^2 e^{-\frac{2k^2}{\Lambda_p^2}}|\vec{R}|^2 (-zk^2+M_h^2+m^2(z-1)+2mM_s-M_s^2)
\\
&+\frac{4}{3} |F^p|^2 e^{-\frac{2k^2}{\Lambda_p^2}} |\vec{R}|^2 \bigg[4\left( \frac{M_h}{2z}-z\frac{k^2+\vec{k}_T^2}{2M_h} \right)^2 +2z \frac{k^2-m^2}{M_h} \left( \frac{M_h}{2z}-z\frac{k^2+\vec{k}_T^2}{2M_h} \right) \bigg] \bigg\}.
\end{aligned}
\end{eqnarray}
As for the twist-2 PDFs $f_1$ and $g_1$, we apply the same spectator model results \cite{Bacchetta:2008af} for uniformity. 
To perform numerical calculation for the $\sin(\phi_h-\phi_R)$ asymmetry in dihadron SIDIS at the COMPASS kinematics, 
we adopt the following kinematical cuts \cite{Sirtl}
\begin{eqnarray}
\begin{aligned}
&\sqrt{s} = 17.4\ \text{GeV} \qquad 0.003 < x < 0.4 \qquad 0.1 < y < 0.9 \qquad 0.2 < z < 0.9 \\
&0.3\text{GeV} < M_h < 1.6\ \text{GeV} \qquad Q^2 > 1\ \text{GeV}^2 \qquad W > 5\ \text{GeV} ,
\end{aligned}
\end{eqnarray}
where $W$ is the invariant mass of photon-nucleon system with $W^2=(P+q)^2 \approx \frac{1-x}{x}Q^2$.
\begin{figure}[htp]
\centering
\includegraphics[height=6.0cm,width=5.7cm]{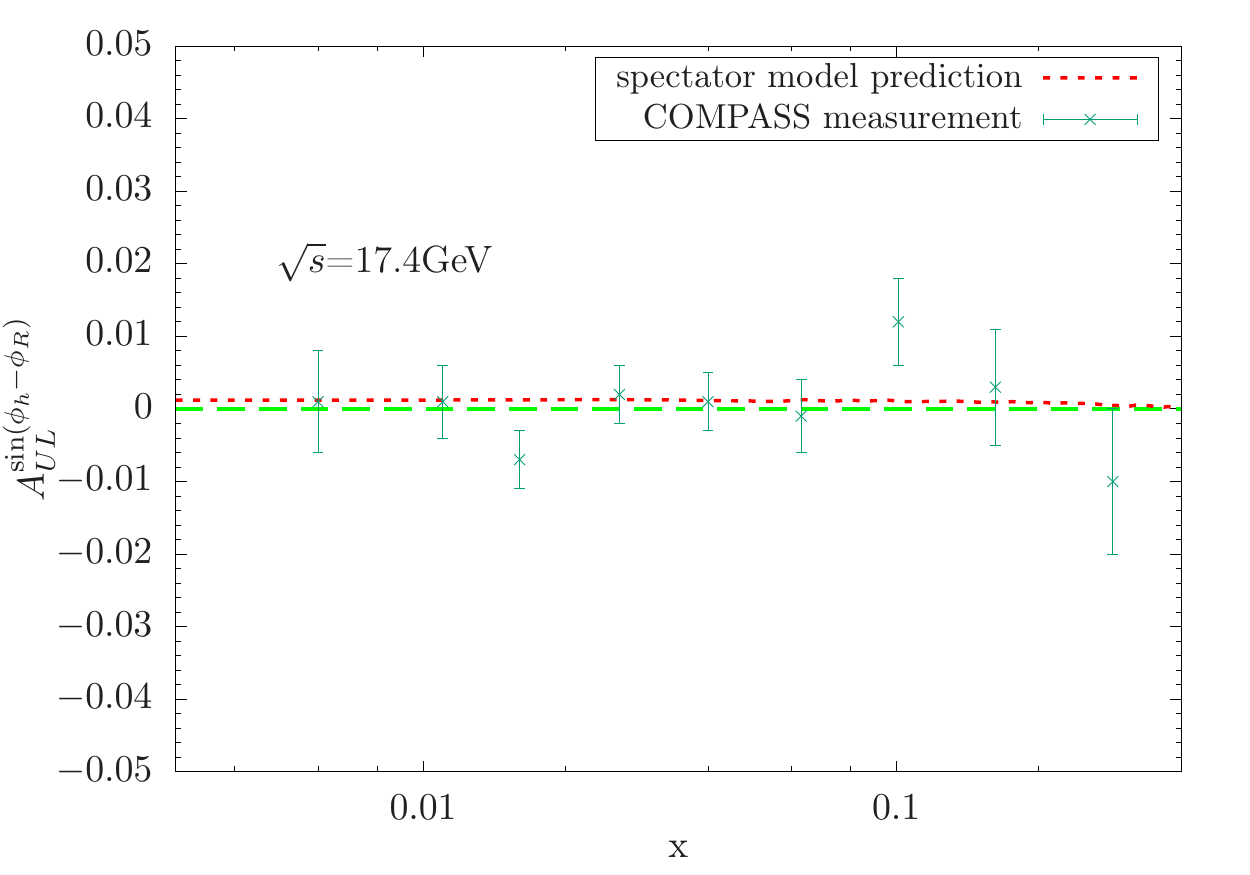}
\includegraphics[height=6.0cm,width=5.7cm]{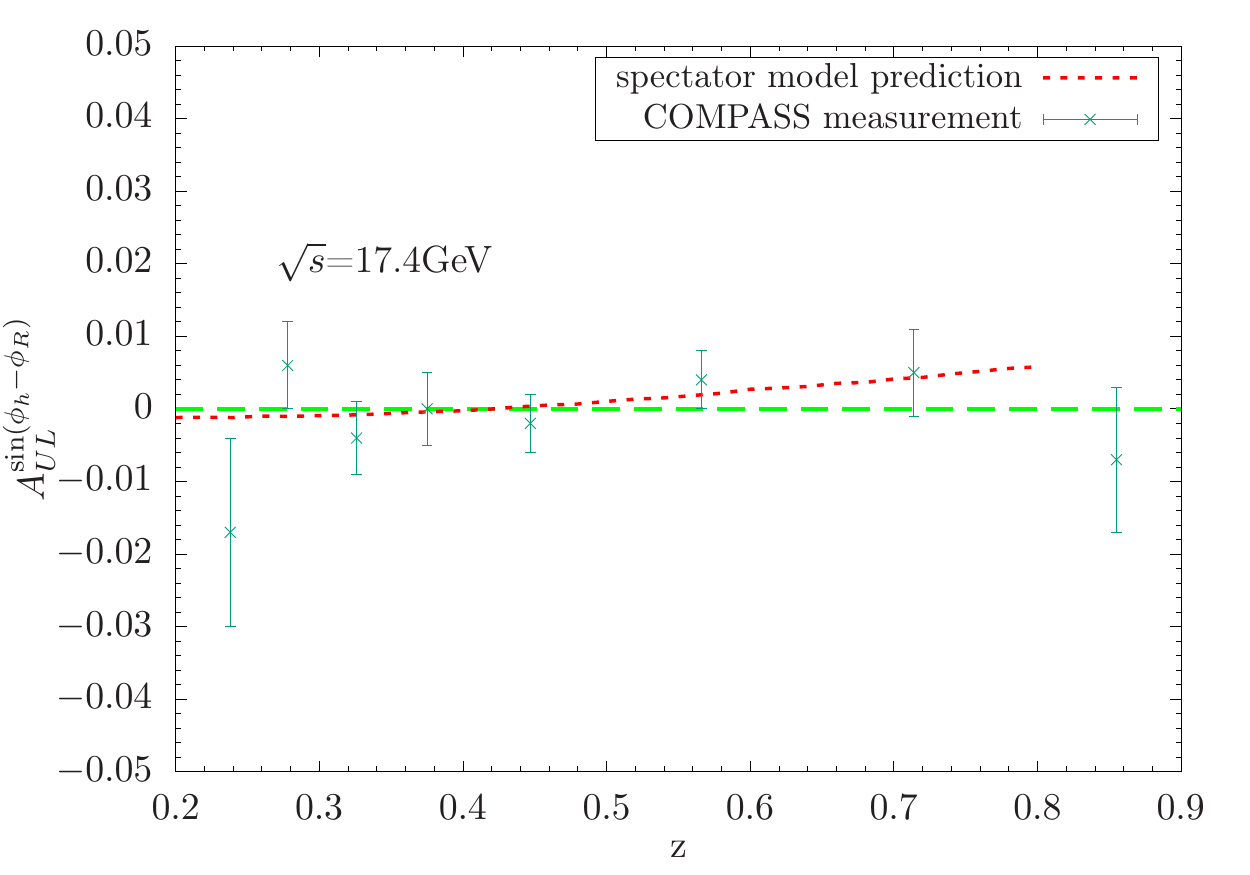}
\includegraphics[height=6.0cm,width=5.7cm]{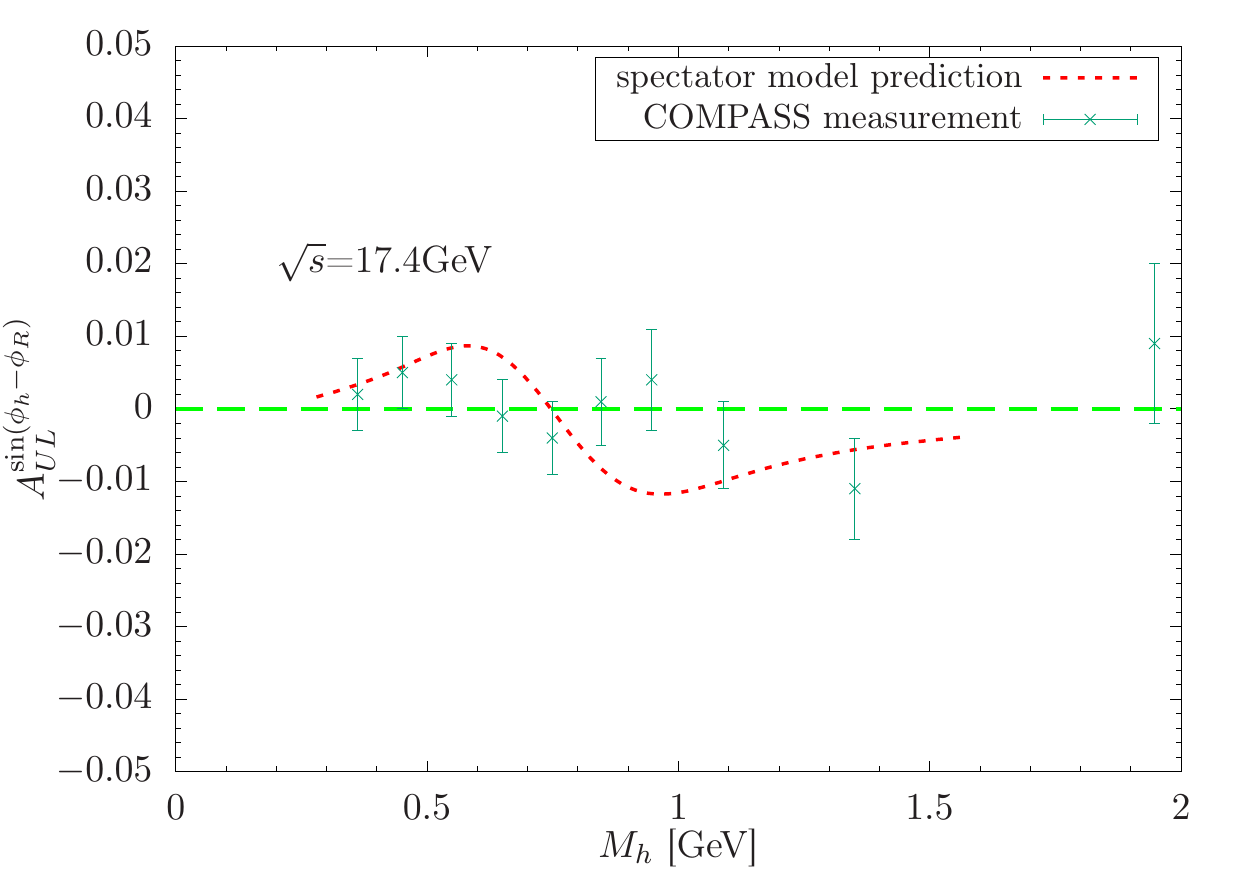}
\caption{\normalsize 
The $\sin(\phi_h-\phi_R)$ azimuthal asymmetry in the SIDIS process of unpolarized muons 
off longitudinally polarized nucleon target as a functions of $x$ (left panel), $z$ (central panel) and $M_h$ (right panel) at COMPASS. 
The full circles with error bars show the preliminary COMPASS data \cite{Sirtl:2017rhi} for comparison. The dashed curves denote the model prediction.}
\label{fig:4}
\end{figure}
Our main results in this work are our predictions for the $\sin(\phi_h-\phi_R)$ 
azimuthal asymmetry in the SIDIS process with unpolarized muons off longitudinally polarized nucleon target, 
as shown in Fig.\ref{fig:4}. The $x$-, $z$- and $M_h$-dependent asymmetries are depicted in the left, 
central and right panels of the figure, respectively. The dashed lines represent our model predictions. 
The full circles with error bars show the preliminary COMPASS data for comparison. 
We can find that the model predictions give a good description of the COMPASS preliminary data being compariable with zero.
\begin{figure}[htp]
\centering
\includegraphics[height=6cm,width=5.7cm]{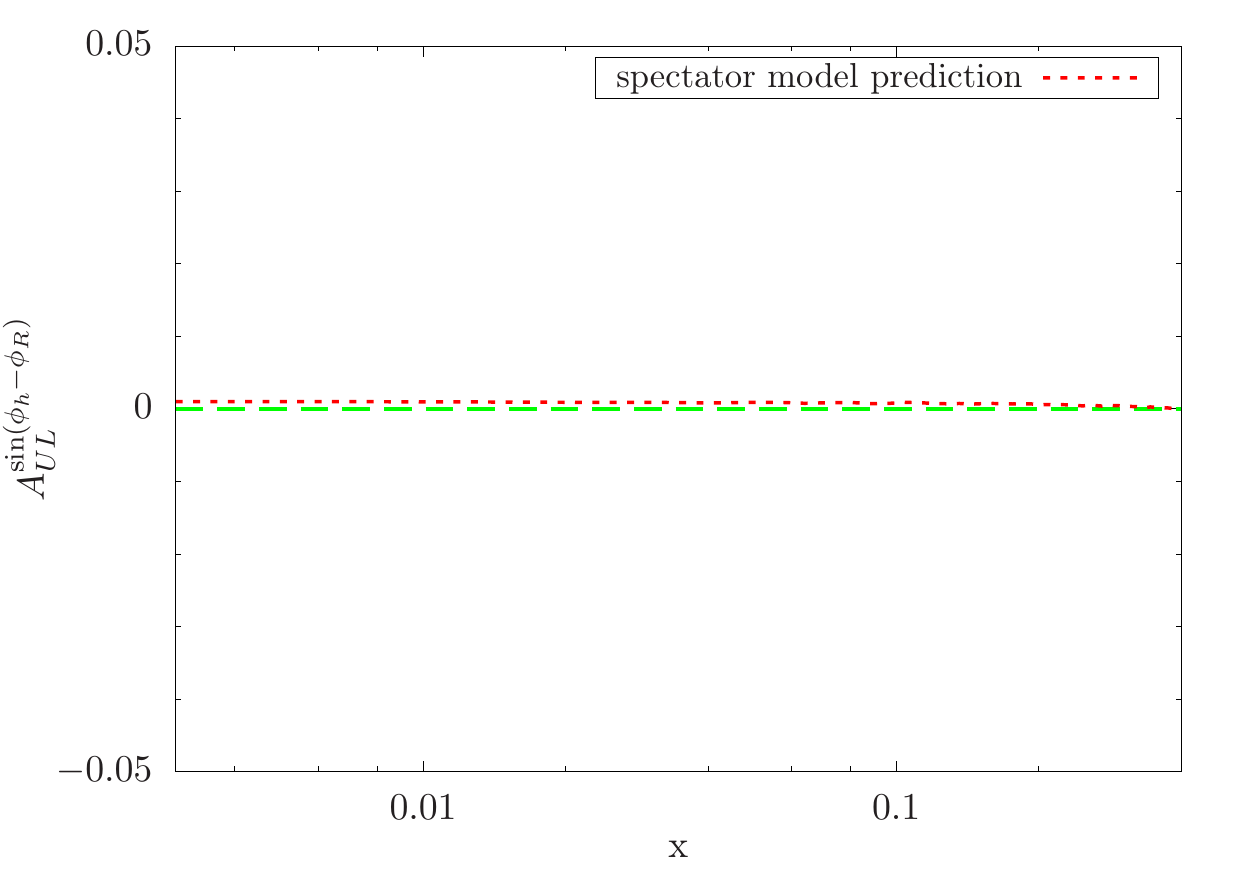}
\includegraphics[height=6cm,width=5.7cm]{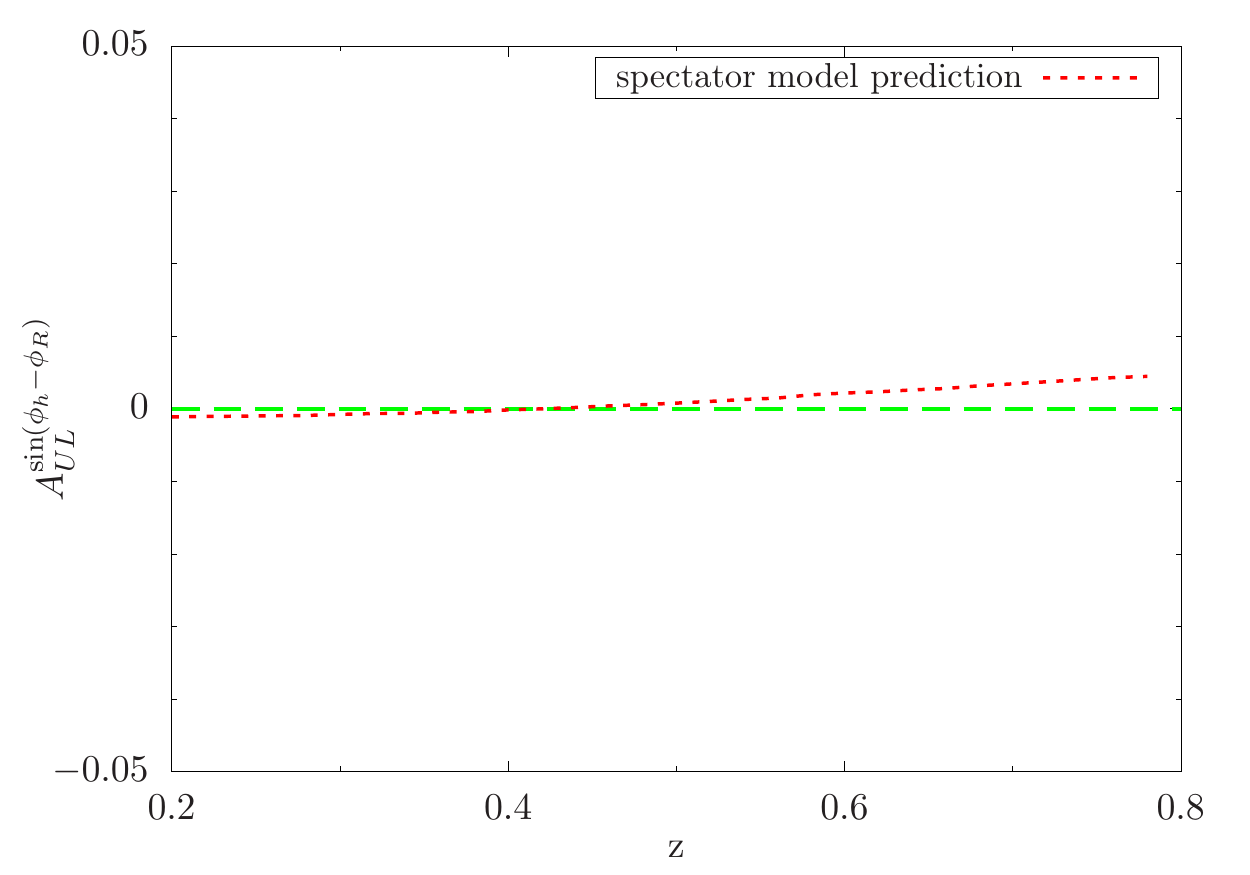}
\includegraphics[height=6cm,width=5.7cm]{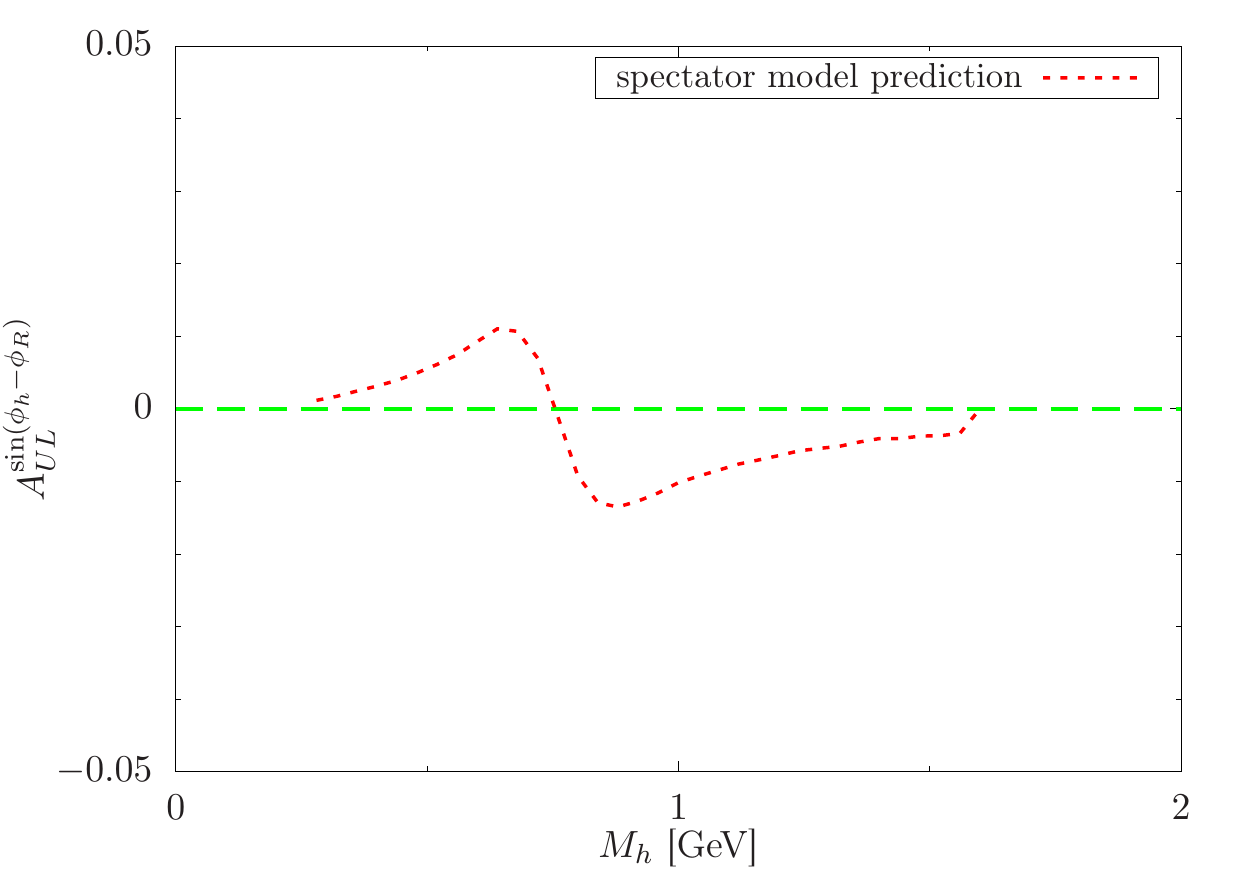}
\caption{\normalsize
The $\sin(\phi_h-\phi_R)$ azimuthal asymmetry in the SIDIS process of unpolarized muons off longitudinally polarized nucleon target 
as a functions of $x$ (left panel), $z$ (central panel) and $M_h$ (right panel) at the EIC ($\sqrt{s}=45$GeV). The dashed curves denote the model prediction.}
\label{fig:5}
\end{figure}

Then, to make a further comparison, we also obtain the $\sin(\phi_h-\phi_R)$ asymmetry at the future EIC. 
Such a facility might be picture-perfect to study this observable. We settle on the following EIC kinematical cuts \cite{Accardi:2012qut}:
\begin{eqnarray}
\begin{aligned}
&\sqrt{s} = 45\ \text{GeV} \qquad 0.001 < x < 0.4 \qquad 0.01 < y < 0.95 \qquad 0.2 < z < 0.8 \\
&0.3\ \text{GeV} < M_h < 1.6\ \text{GeV} \qquad Q^2 > 1\ \text{GeV}^2 \qquad W^2 > 10\ \text{GeV}^2 .
\end{aligned}
\end{eqnarray}
The $x$-, $z$- and $M_h$-dependent asymmetries are plotted in the left, central, and right panels in Fig.\ref{fig:5}. 
We find that the overall tendency of the asymmetry at the EIC is similar to that at COMPASS. 
Since the size of the asymmetry is lightly smaller than that at COMPASS, the results are still compariable with zero at the kinematics of EIC.
It is also possible to conceive a set of kinematical cuts reach for the future EIC 
but not for COMPASS to observe if the future EIC could bring new information on asymmetries.
We can reach the following EIC kinematical cuts for a future EIC:
\begin{eqnarray}
\begin{aligned}
&\sqrt{s} = 100\ \text{GeV} \qquad 0.001 < x < 0.7 \qquad 0.01 < y < 0.95 \qquad 0.2 < z < 0.8 \\
&0.3\ \text{GeV} < M_h < 1.6\ \text{GeV} \qquad Q^2 > 25 \text{GeV}^2 \qquad W > 10\ \text{GeV} .
\end{aligned}
\end{eqnarray}
\begin{figure}[htp]
\centering
\includegraphics[height=6cm,width=5.7cm]{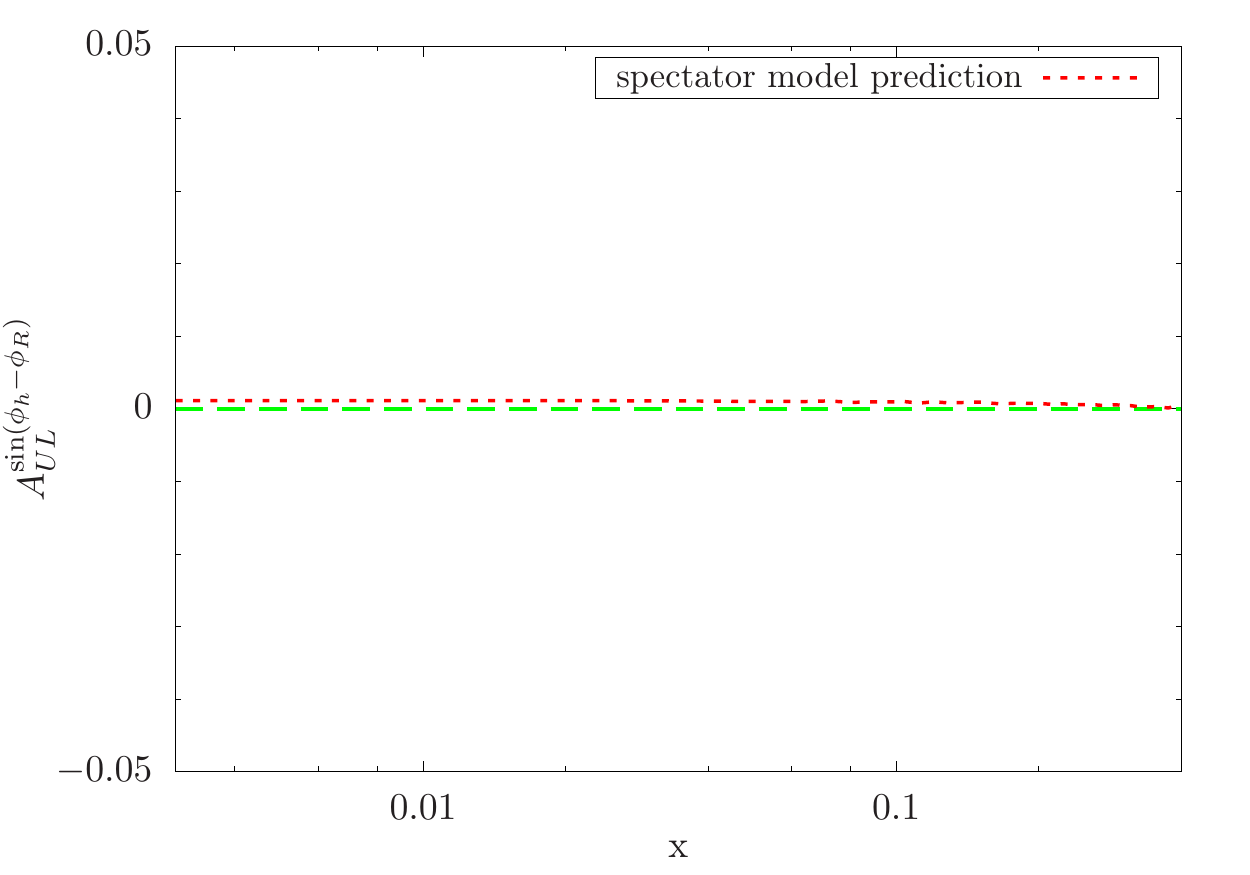}
\includegraphics[height=6cm,width=5.7cm]{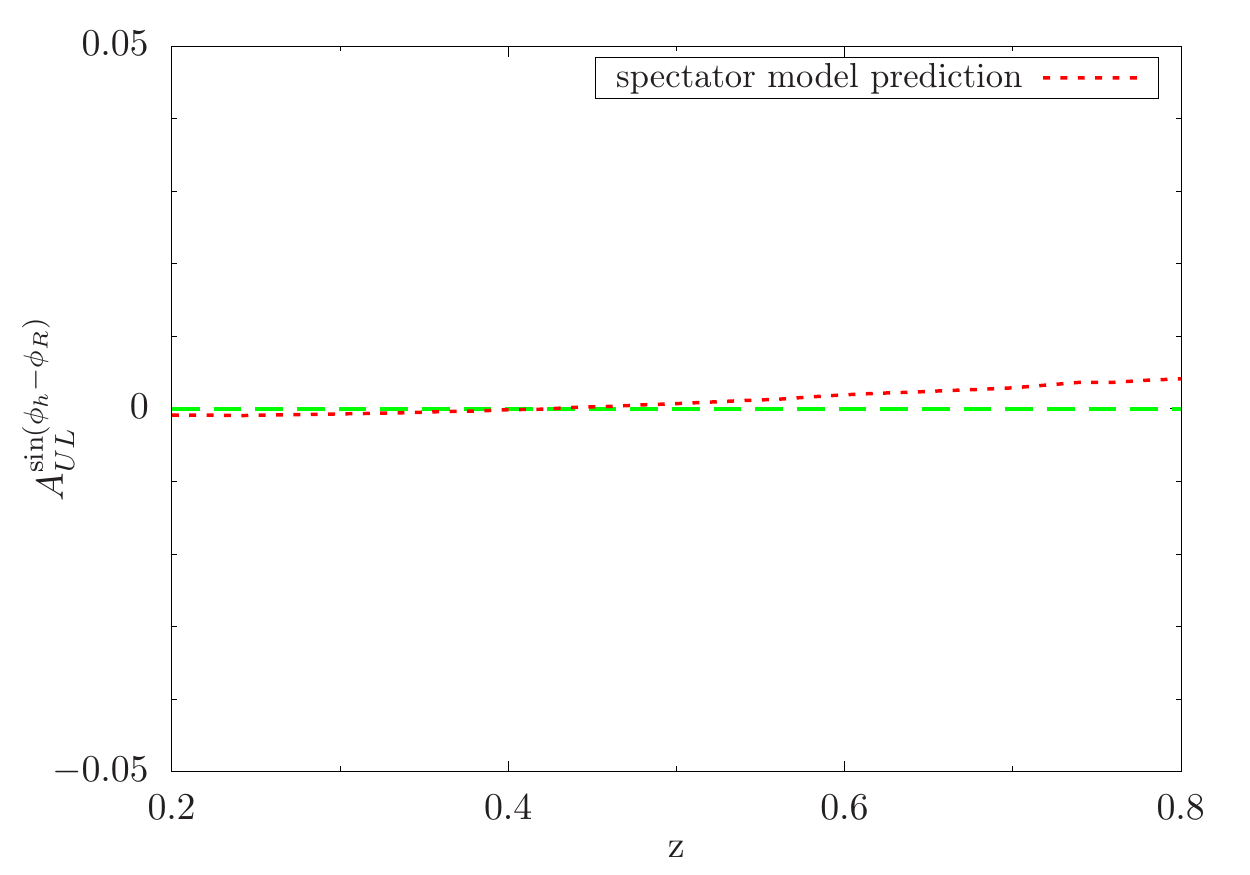}
\includegraphics[height=6cm,width=5.7cm]{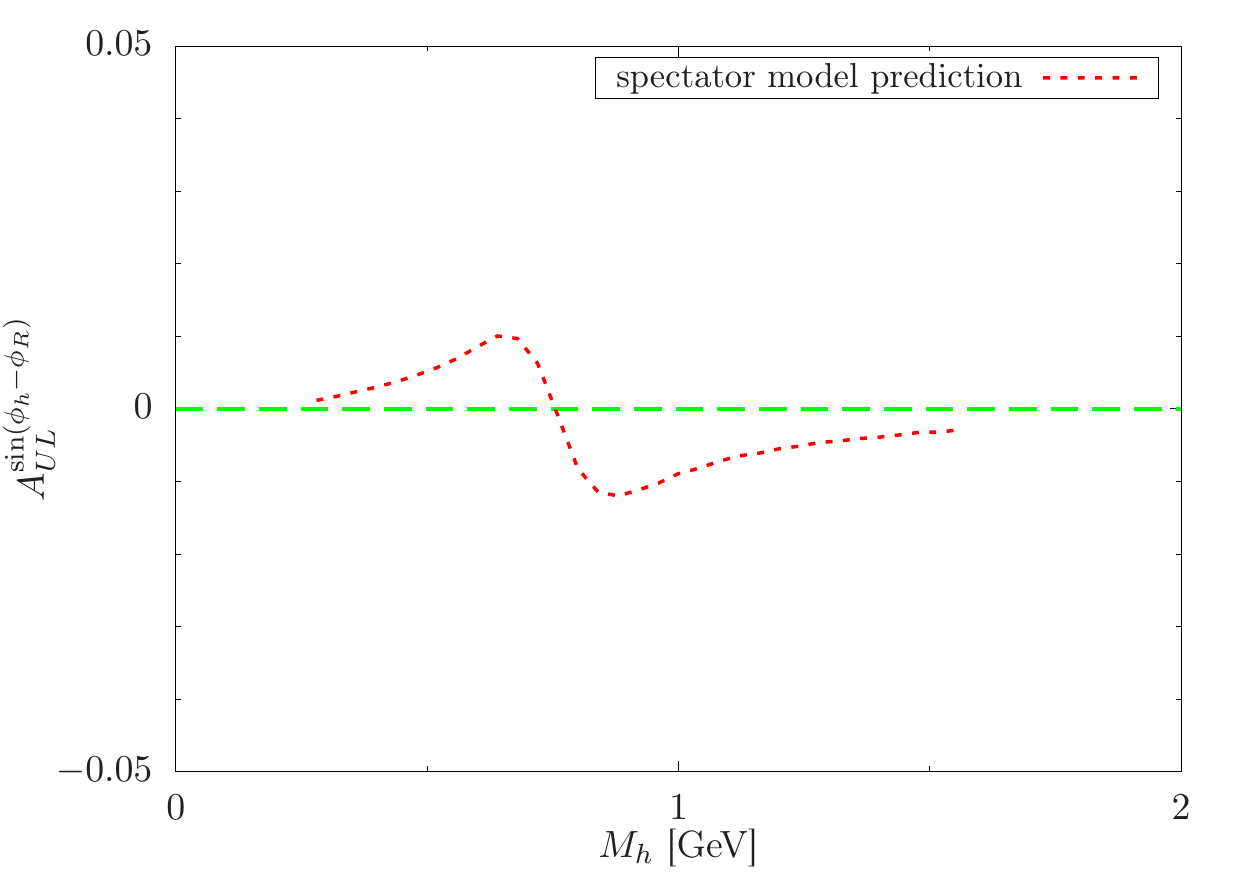}
\caption{\normalsize
The $\sin(\phi_h-\phi_R)$ azimuthal asymmetry in the SIDIS process of unpolarized muons 
off longitudinally polarized nucleon target as a functions of $x$ (left panel), 
$z$ (central panel) and $M_h$ (right panel) at the future EIC ($\sqrt{s}=100$GeV). The dashed curves denote the model prediction.}
\label{fig:6}
\end{figure}
The $x$-, $z$- and $M_h$-dependent asymmetries are plotted in the left, central, and right panels in Fig.\ref{fig:6}.
We find that the asymmeties are not prompted by the future EIC kinematics.

In addition, the available theoretical bounds on $G_1^\perp$(see Eq.(55) of \cite{Radici:2001na}) would give a much larger asymmetry, and therefore the COMPASS data and the present predictions strongly constrain this DiFF within their uncertainty.

Finally, the expected size of $A_{UL}^{\sin(\phi_h-\phi_R)}$ 
in the NJL quark model \cite{Matevosyan:2013aka,Matevosyan:2013eia,Matevosyan:2017alv,Matevosyan:2017uls} is zero, 
which is basically consistent with the spectator model result. 
This conclusion can be reached by considering Eq.(12) of Ref. \cite{Matevosyan:2017liq} 
together with both Eq.(19) and Eq.(26) in Ref. \cite{Matevosyan:2017uls}.
In addition, Ref. \cite{Matevosyan:2017liq} proposed a measurement 
which will permit us to access the helicity-dependent DiFF in forward two-hadron production in SIDIS. 
We will make a prediction for this measurement
in a upcoming work using the spectator model result of helicity-dependent DiFFs.

\section{Conclusion}
\label{VI}

In this work, we have considered the single spin asymmetry with a $\sin(\phi_h-\phi_R)$ modulation of dihadron production in SIDIS. 
With the spectator model result for $D_{1,OO}$ at hand, we worked out the T-odd DiFF $G_{1,OT}^\perp$ 
by considering the real and imaginary loop contributions. Using the partial wave expansion, 
we found that $G_{1,OT}^\perp$ comes from the interference contribution of the $s$- and $p$-waves. 
By using the numerical results of the DiFFs and PDFs, we present the prediction for the $\sin(\phi_h-\phi_R)$ asymmetry 
and compare it with the COMPASS measurements. Our result yields a good description of the vanished COMPASS data.

\begin{acknowledgments}
Xuan Luo thanks professor Alessandro Bacchetta, Marco Radici and Aram Kotzinian for their patient guidance. 
The authors also thank professor Zhun Lu and doctor Yongliang Yang for useful discussions. 
Hao Sun is supported by the National Natural Science Foundation of China (Grant No.11675033) 
and by the Fundamental Research Funds for the Central Universities (Grant No. DUT18LK27).
\end{acknowledgments}

\bibliography{v1}
\end{document}